\documentclass[12pt]{iopart}
\usepackage{tikz}
\usepackage{graphicx}
\graphicspath{{Plots/}}
\usepackage{float}% to force figure placement 
\usepackage{bm}% bold math
\usepackage[hidelinks]{hyperref}
\usepackage{slashed}
\newcommand{\spindersphere}{{}_s\slashed{\Delta}}
\usepackage{xcolor}

\begin{document}

\title[Teukolsky QNMs and QNEs in HPHC coordinates]{Computing the
quasinormal modes and eigenfunctions  
for the Teukolsky equation
using horizon penetrating, hyperboloidally compactified coordinates
}

\author{Justin L. Ripley}

\address{DAMTP,
Centre for Mathematical Sciences,
University of Cambridge,
Wilberforce Road, Cambridge CB3 0WA, UK.}
\ead{jr860@cam.ac.uk}
\vspace{10pt}
\begin{indented}
\item[]June 2022
\end{indented}

%%%%%%%%%%%%%%%%%%%%%%%%%%%%%%%%%%%%%%%%%%%%%%%%%%%%%%%%%%%%%%%%%%%%%%%%%%%%%%

\begin{abstract}
   We study the quasinormal mode eigenvalues and eigenfunctions for
   the Teukolsky equation in a horizon penetrating,
   hyperboloidally compactified (HPHC) coordinate system. 
   Following earlier work by Zengino\u{g}lu \cite{zenginouglu2011geometric}, 
   we show that the quasinormal eigenfunctions for the 
   Teukolsky equation are regular from the black hole horizon to 
   future null infinity in these coordinates. 
   We then present several example quasinormal eigenfunction 
   solutions, and study some of
   their properties in the near-extremal Kerr limit.
\end{abstract}
\noindent{\emph{keywords}: 
   black holes, quasinormal modes, quasinormal eigenfunctions, 
   hyperboloidal compactification
}
%\submitto{\CQG}
\maketitle
%%%%%%%%%%%%%%%%%%%%%%%%%%%%%%%%%%%%%%%%%%%%%%%%%%%%%%%%%%%%%%%%%%%%%%%%%%%%%%

%%%%%%%%%%%%%%%%%%%%%%%%%%%%%%%%%%%%%%%%%%%%%%%%%%%%%%%%%%%%%%%%%%%%%%%%%%%%%%
\section{Introduction}
%%%%%%%%%%%%%%%%%%%%%%%%%%%%%%%%%%%%%%%%%%%%%%%%%%%%%%%%%%%%%%%%%%%%%%%%%%%%%%

The Teukolsky equation describes the dynamics of linear spin $s$ fields 
on the Kerr black hole spacetime \cite{Teukolsky:1973ha}.
As this equation describes an inherently dissipative system (waves can fall
into the black hole, or propagate to future null infinity), the Teukolsky
equation does not have mode solutions, but instead has
\emph{quasi}normal mode solutions.
In this work we will be interested in computing not just the
quasinormal modes (QNMs) of the Teukolsky equation, 
but also the quasinormal \emph{eigenfunction} (QNEs)
associated with each mode.

\begin{figure}
   \centering
   \begin{tikzpicture}
      \draw [very thick] 
         (0,0) -- (2.5,2.5) -- (0,5) -- (-2.5,2.5) -- (0,0);
      \draw [gray,dashed,very thick] 
         (-2,3) .. controls (-1,2.2) and (1.5,2) .. (2,3);
      \draw [gray,dashed,very thick] 
         (-1.4,3.6) .. controls (-0.3,3) and (0.8,2.8) .. (1.4,3.6);
      \draw [gray,dashed,very thick] 
         (-0.8,4.2) .. controls (-0.2,3.7) and (0.4,3.6) .. (0.8,4.2);
      \draw [gray,dotted,very thick] 
         (-2.5,2.5) .. controls (-0.8,3.8) and (0.8,3.8) .. (2.5,2.5);
      \draw [gray,dotted,very thick] 
         (-2.5,2.5) .. controls (-0.8,1.2) and (0.8,1.2) .. (2.5,2.5);
      \draw [gray,dotted,very thick] 
         (-2.5,2.5) -- (2.5,2.5);
      \node at (2,3.8) {$\mathcal{J}^+$};
      \node at (-1.8,3.8) {$\mathcal{H}^+$};
      \node at (0.05,5.3) {$i^+$};
      \node at (2.8,2.5) {$i^0$};
      \node at (-2.8,2.5) {$\mathcal{B}$};
   \end{tikzpicture}
   \caption{Schematic Penrose diagram of the Kerr black hole spacetime
      exterior to the future horizon
      that illustrates the horizon-penetrating, hyperboloidally compactified 
      coordinates used in this article. 
      The dotted lines describe 
      $t=const.$ hypersurfaces in Boyer-Lindquist coordinates,
      while the dashed lines describe $\tau=const.$ hypersurfaces in
      the horizon-penetrating, hyperboloidally compactified coordinates we use.
      Here $\mathcal{H}^+$ is the future black hole horizon, $\mathcal{J}^+$
      is future null infinity, $i^+$ is future timelike infinity,
      $i^0$ is spacelike infinity, and $\mathcal{B}$ is the bifurcation sphere
      of the black hole.}
      \label{fig:penrose_diagram_bh}
\end{figure}
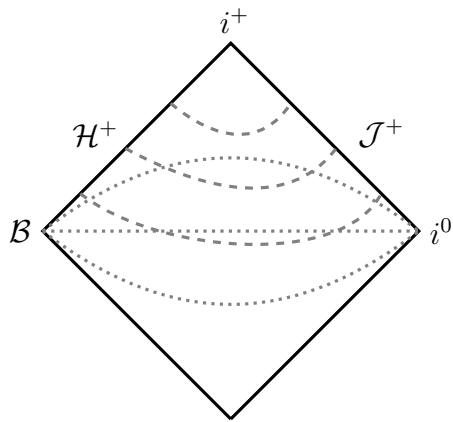

The QNMs of the Teukolsky equation 
have found use in astrophysics, theoretical
physics, and mathematical relativity 
\cite{Nollert:1999ji,Kokkotas:1999bd,Berti:2009kk,Konoplya:2011qq}. 
Teukolsky QNM mode calculations are typically computed using 
coordinates where the constant time hypersurfaces intersect
the bifurcation sphere and spatial infinity
(for example Boyer-Lindquist coordinates have this 
property \cite{doi:10.1063/1.1705193})
\cite{
   Teukolsky:1973ha,PhysRev.108.1063,PhysRevLett.24.737,doi:10.1063/1.1666175,
   Leaver:1985ax}.
As was pointed out though by Zengino\u{g}lu,
horizon-penetrating, hyperboloidally compactified (HPHC) 
coordinates--that is, 
coordinates where constant time hypersurfaces intersect both 
the black hole horizon and future null infinity 
(see Fig.~\ref{fig:penrose_diagram_bh})--can 
be considered a more ``natural'' set of coordinates 
to study black hole perturbations \cite{zenginouglu2011geometric}.
This is because in HPHC coordinates, on constant time
hypersurfaces the QNEs 
are regular at the horizon and 
at future null infinity\footnote{Press and Teukolsky 
were the first to note that the QNEs 
of the Teukolsky equation
do not blow up at the black hole horizon when one works with an appropriate
tetrad in horizon penetrating coordinates, and that the QNEs 
do not blow up at future null
infinity if one works in outgoing coordinates \cite{Teukolsky:1974yv}.}.
By contrast, on constant time hypersurfaces the QNEs blow up exponentially at
the bifurcation sphere and at spatial infinity.
Moreover, a timelike observer can never reach the bifurcation sphere
or spatial infinity, so from a physical perspective it is not
necessary to know how the QNEs behave near those two locations.
Here we extend and complete the calculations 
begun in \cite{zenginouglu2011geometric},
by computing the QNMs and QNEs 
for the Kerr black hole in HPHC coordinates\footnote{In contrast to
calculations of QNEs, which are computed in the frequency
domain, there has several works that compute the
evolution of the Teukolsky equation in the time domain using
HPHC coordinates \cite{Zenginoglu:2011zz,Harms:2013ib,
Csukas:2019kcb,Ripley:2020xby}.}
We use a spectral/pseudospectral method to compute the QNEs.

Recently the \emph{pseudospectrum} of the QNMs of
Schwarzschild black holes \cite{Jaramillo:2020tuu} and
Reissner-Nordstrom black holes \cite{Destounis:2021lum}
were computed in HPHC coordinates.
The pseudospectrum of a mode solution roughly
captures how sensitive the solution is to perturbations of the
underlying equation of motion \cite{trefethen2005spectra}. 
In this sense, the pseudospectrum of the Teukolsky
equation then quantifies the stability of QNMs to perturbations of
the underlying Kerr spacetime\footnote{For other recent 
attempts to investigate the stability of the
quasinormal mode solutions to the Teukolsky equation; see 
\cite{Loutrel:2020wbw,Ripley:2020xby,Cheung:2021bol,Sberna:2021eui}.}.
Computing the pseudospectrum requires evaluating 
QNEs in a suitable norm, which is one reason why HPHC coordinates were
used in \cite{Jaramillo:2020tuu,Destounis:2021lum}: in these
coordinates the QNEs remain finite over
the entire exterior of the black hole, 
which makes finding a well-behaved norm relatively straightforward.
This work is partly motivated by the pseudospectral
research program initiated in \cite{Jaramillo:2020tuu,Destounis:2021lum},
as (to our knowledge) no equivalent computation of the 
QNEs of the Teukolsky equation in HPHC coordinates has 
yet been completed.

This note is organized as follows.
We first derive the Teukolsky equation in HPHC coordinates,
and separate the resulting equation into two ordinary differential 
equations (ODEs).
We use of a spectral method to discretize the angular equation,
and a pseudospectral method to discretize the radial equation.
We then present a method to numerically compute 
the QNMs and QNEs of the Teukolsky equation
by rephrasing the two discretized ODEs as a joint 
eigenvalue problem.
We present some example QNE solutions, and study
their functional form in the (near) extremal limit $a\to M$.
In the appendices we provide a code comparison of our code
to the \texttt{qnm} code \cite{Stein:2019mop}, 
present a convergence study of an example QNE solution,
and review some properties of orthogonal polynomials
which we make use of in our (pseudo)spectral code.

The metric signature is $-+++$, and we set $G=c=\hbar=1$.
The real and imaginary part of a number are denoted by
$\mathcal{R}$ and $\mathcal{I}$, respectively. 

Our code is available online \cite{code_online}.

%%%%%%%%%%%%%%%%%%%%%%%%%%%%%%%%%%%%%%%%%%%%%%%%%%%%%%%%%%%%%%%%%%%%%%%%%%%%%%
\section{Hyperboloidal compactification of the Teukolsky equation
   \label{sec:hyperboloidal_compactification}
}
We first rewrite the Teukolsky equation in a HPHC coordinate system.
Our general approach follows \cite{Ripley:2020xby}; see also 
\cite{zenginouglu2011geometric,
zenginouglu2008hyperboloidal,
PanossoMacedo:2019npm} 
for other potential choices of HPHC coordinates.
The Teukolsky equation in Boyer-Lindquist coordinates is
\cite{Teukolsky:1973ha}
\begin{eqnarray}
   \fl
   \left[
      \frac{\left(r^2+a^2\right)^2}{\Delta}
      -
      a^2\sin^2\theta
   \right]
   \partial_t^2\psi
   -
   \Delta^{-s}\partial_r\left(\Delta^{s+1}\partial_r\psi\right)
   +
   \frac{4Mar}{\Delta}\partial_t\partial_{\varphi}\psi
   +
   \frac{a^2}{\Delta}
   \partial_{\varphi}^2\psi
   -
   \spindersphere\psi
   \nonumber\\
   -
   2s\frac{a(r-M)}{\Delta}\partial_{\varphi}\psi
   -
   2s
   \left[
      \frac{M\left(r^2-a^2\right)}{\Delta}
      -
      r
      -
      ia\cos\theta
   \right]
   \partial_t\psi
   =
   0
   ,
\end{eqnarray}
where $M$ is the black hole mass, $a$ is the black hole spin,
and $s$ is the spin of the wave ($\pm2$ for gravity, $\pm1$ for
electromagnetism, etc).
The spin-weighted spherical Laplacian is
\begin{equation}
   \spindersphere\psi
   \equiv
   \frac{1}{\sin\theta}\partial_{\theta}\left(
      \sin\theta\partial_{\theta}\psi
   \right)
   +
   \left(
      s
      -
      \frac{\left(-i\partial_{\varphi}+s\cos\theta\right)^2}{\sin^2\theta}
   \right)
   \psi
   .
\end{equation}
We also have used the standard notation
\begin{equation}
   \Delta
   \equiv
   r^2-2Mr+a^2
   .
\end{equation}
The zeros of $\Delta$ determine the location of the inner and outer horizons:
\begin{equation}
   r_{\pm}
   \equiv
   M\pm\sqrt{M^2-a^2}
   .
\end{equation}

We transform to ingoing coordinates by defining the variables 
\begin{eqnarray}
   dv
   \equiv
   dt
   +
   \frac{2Mr}{\Delta}dr
   ,\qquad
   d\phi
   \equiv
   d\varphi
   +
   \frac{a}{\Delta}dr
   .
\end{eqnarray}
We also radially rescale $\psi$ to make the Teukolsky equation
regular at the horizon, and to remove the ``long-range potential''
in the Teukolsky equation \cite{SASAKI198268,Hughes:2000pf}
\begin{equation}
   \psi
   \equiv
   \frac{1}{r}\Delta^{-s} \Psi
   .
\end{equation}
The Teukolsky equation now reads
\begin{eqnarray}
   \fl
   \left(
      r^2
      +
      2Mr
      +
      a^2\cos^2\theta
   \right)
   \partial_v^2\Psi
   -
   4Mr\partial_v\partial_r\Psi
   -
   \Delta\partial_r^2\Psi
   -
   2a\partial_r\partial_{\phi}\Psi
   -
   \spindersphere\Psi
   \nonumber
   \\
   \fl
   +
   2
   \left[
      M
      +
      s\left(
         M
         +
         r
         +
         ia\cos\theta
      \right)
   \right]
   \partial_v\Psi
   +
   2\left[
      -
      M
      +
      \frac{a^2}{r}
      +
      s\left(r-M\right)
   \right]
   \partial_r\Psi
   +
   \frac{2a}{r}
   \partial_{\phi}\Psi
   \nonumber
   \\
   \fl
   +
   2\left(
      \frac{sM}{r}
      +
      \frac{Mr-a^2}{r^2}
   \right)
   \Psi
   =
   0
   .
\end{eqnarray}

We next transform the time variable to achieve a hyperboloidal
slicing of the spacetime; this will make
the Teukolsky equation regular at future null infinity
\cite{zenginouglu2008hyperboloidal,zenginouglu2011geometric}.
We define the hyperboloidal time variable $\tau$ 
\begin{equation}
   d\tau
   \equiv
   dv
   +
   \frac{dh}{dr}dr
   ,
\end{equation}
where $h(r)$ is a ``height'' function designed so that the radially ingoing
characteristic speed is zero at $r=\infty$. Ultimately we find
\begin{equation}
   \frac{dh}{dr}
   =
   -
   1
   -
   \frac{4M}{r}
   ,
\end{equation}
to be a suitable height function
\cite{zenginouglu2008hyperboloidal,PanossoMacedo:2019npm,Ripley:2020xby}.
We choose the radial compactification
\begin{equation}
   \rho
   \equiv
   \frac{1}{r}
   ,
\end{equation}
so $r=\infty$ is located at $\rho=0$.
The Teukolsky equation now reads
\begin{eqnarray}
\label{eq:hyperboloidal_teuk_eqn}
   \fl
   \left[
      16M^2
      -
      a^2\sin^2\theta
      +
      8M\left(4M^2-a^2\right)\rho
      -
      16a^2M^2\rho^2
   \right]
   \partial_{\tau}^2\Psi
   -
   \rho^4\Delta\partial_{\rho}^2\Psi
   -
   \spindersphere\Psi
   \nonumber
   \\
   \fl
   -
   2\left[
      1
      +
      \left(a^2 - 8M^2\right)\rho^2
      +
      4a^2M\rho^3
   \right]
   \partial_{\tau}\partial_{\rho}\Psi
   +
   2a\rho^2\partial_{\rho}\partial_{\phi}\Psi
   +
   2a\left(1+4M\rho\right)\partial_{\tau}\partial_{\phi}\Psi
   \nonumber
   \\
   \fl
   +
   2
   \left[
      s\left(
         -
         2M
         +
         ia\cos\theta
      \right)
      +
      \left(
         4M^2\left\{s+2\right\}
         -
         a^2
      \right)
      \rho
      -
      6Ma^2\rho^2
   \right]
   \partial_{\tau}\Psi
   \nonumber
   \\
   \fl
   +
   2
   \left[
      -
      1
      -
      s
      +
      \left(s+3\right)M\rho
      -
      2a^2\rho^2
   \right]
   \rho
   \partial_{\rho}\Psi
   +
   2a\rho\partial_{\phi}\Psi
   \nonumber
   \\
   \fl
   +
   2\left(
      Ms
      +
      M
      -
      a^2\rho
   \right)
   \rho
   \Psi
   =
   0
   .
\end{eqnarray}
The Teukolsky equation remains regular at the radial endpoints
$\rho=0,\rho=\rho_+$ (although it is still singular at $\theta=0,\pi$). 
At future null infinity (located at $\rho=0$), the radial
ingoing characteristic speed is zero.

For reference, and to verify that $\rho=0$ limits to an asymptotically
flat spacetime, we next present the Kerr line element in the
coordinates $\{\tau,r\equiv1/\rho,\theta,\phi\}$.
Defining $\Sigma\equiv r^2+a^2\cos^2\theta$, the line element is:
\begin{eqnarray}
   \label{eq:kerr_line_element}
   \fl
   ds^2
   =
   -
   \left(
      1
      -
      \frac{2Mr}{\Sigma}
   \right)d\tau^2
   +
   8M
   \frac{1}{\Sigma}
   \left(
      1
      +
      \frac{2M}{r}
   \right)
   \left(
      2M
      -
      \frac{a^2}{r}\cos^2\theta
   \right)dr^2 
   \nonumber\\
   \fl
   -
   2\left(
      1
      +
      \frac{4M}{r}
      -
      \frac{4M\left(r+2M\right)}{\Sigma}
   \right)d\tau dr
   -
   4a\frac{Mr}{\Sigma}\sin^2\theta d\tau d\phi
   \nonumber\\
   \fl
   -
   2a\left(
      1
      +
      4M\frac{r+2M}{\Sigma}
   \right)
   \sin^2\theta drd\phi
   +
   \Sigma d\theta^2
   +
   \left(a^2+r^2+2Mr\frac{a^2}{\Sigma}\sin^2\theta\right)
   \sin^2\theta
   d\phi^2
   .
\end{eqnarray}
Holding $\tau=const.,\theta=const.,$ and $\phi=const.$, 
the radial proper distance line element $dr_p$ is
\begin{equation}
   \label{eq:kerr_radial_proper_distance}
   dr_p
   =
   8M\frac{1}{\Sigma}
   \left(1+\frac{2M}{r}\right)
   \left(2M-\frac{a^2}{r}\cos^2\theta\right)
   dr
   .
\end{equation}
This is finite and bounded for all $r_+<r<\infty$, thus we see that the
proper radial distance from any point $r_+<r_0<\infty$
to the black hole horizon is always bounded, 
including in the extremal limit $a/M\to1$.
In the limit $r\to\infty$, the metric is flat, including
in the extremal limit \cite{zenginouglu2008hyperboloidal,PanossoMacedo:2019npm}
\begin{equation}
   \fl
   \lim_{r\to\infty}ds^2
   =
   -
   d\tau^2
   -
   2d\tau dr
   +
   r^2\left(
      d\theta^2 
      + 
      \sin^2\theta d\phi^2 
      + 
      \mathcal{O}\left(\frac{1}{r^2}\right)
   \right)
   +
   \mathcal{O}\left(1\right)dr d\phi
   .
\end{equation}
To summarize:
we see that we have chosen coordinates so that on $\tau=const.$
hypersurfaces, even in the limit $a/M\to 1$,
$r=\infty$ $(\rho=0)$ corresponds to future null infinity,
$r=r_+$ $(\rho=\rho_+)$ corresponds to the black hole horizon,
and the proper radial distance to the black hole horizon remains bounded.

%%%%%%%%%%%%%%%%%%%%%%%%%%%%%%%%%%%%%%%%%%%%%%%%%%%%%%%%%%%%%%%%%%%%%%%%%%%%%%
\section{Quasinormal mode solutions to the Teukolsky equation
   \label{sec:quasinormal_mode_solutions}
}
%-----------------------------------------------------------------------------
\subsection{Separating the Teukolsky equation into ODEs
   \label{sec:seperating_teuk}
}
To determine the quasinormal modes and eigenfunctions
of the Teukolsky equation, 
we first apply separation of variables \cite{Teukolsky:1973ha}
\begin{equation}
   \Psi\left(\tau,\rho,\theta,\phi\right)
   =
   e^{-i\omega \tau + i m\phi}R(\rho)S(\theta)
   .
\end{equation}
With this, the Teukolsky equation separates into 
a radial and an angular equation:
\begin{eqnarray}
\label{eq:radial_rho}
   \fl
   -
   \rho^2\hat{\Delta}\left(\rho\right)
   \frac{d^2R}{d\rho^2}
   +
   A\left(\omega,m,\rho\right)\frac{dR}{d\rho}
   +
   \left(
      B\left(\omega,m,\rho\right)
      -
      {}_s\Lambda^m_l
   \right)
   R
   =
   0
   ,\\
   \label{eq:spheroidal_laplacian}
   \fl
   \frac{1}{\sin\theta}
   \frac{d}{d\theta}\left(\sin\theta\frac{dS}{d\theta}\right)
   +
   \left(
      s
      -
      \frac{\left(m+s\cos\theta\right)^2}{\sin^2\theta}
      -
      2a\omega s\cos\theta
      +
      a^2\omega^2\cos^2\theta
      +
      {}_s\Lambda^m_l
   \right)
   S
   =
   0
   .
\end{eqnarray}
We have defined
\begin{eqnarray}
\label{eq:definitions_radial_teukolsky}
   \fl
   \hat{\Delta}\left(\rho\right)
   \equiv
   1-2M\rho+a^2\rho^2
   ,\\
   \fl
   A\left(\omega,m,\rho\right)
   \equiv
      2i\omega
      -
      2\left(1+s\right)\rho
      +
      2\left[
         i\omega\left(a^2-8M^2\right)
         +
         ima
         +
         \left(s+3\right)M
      \right]\rho^2
      \nonumber\\
      \fl
      +
      4\left[2i\omega M-1\right]a^2\rho^3
   ,\\
   \fl
   B\left(\omega,m,\rho\right)
   \equiv
      \left(a^2-16M^2\right)\omega^2
      +
      2\left(ma+2isM\right)\omega
      \nonumber\\
      \fl
      +
      2\left[
         4\left(a^2-4M^2\right)M\omega^2
         +
         \left(4m aM - 4i\left(s+2\right)M^2 + ia^2\right)\omega
         +
         ima
         +
         \left(s+1\right)M
      \right]
      \rho
      \nonumber\\
      \fl
      +
      2\left(
         8M^2\omega^2
         +
         6iM\omega
         -
         1
      \right)
      a^2\rho^2
   .
\end{eqnarray}
The separation constant ${}_s\Lambda^m_l$ can be thought of as a
function of $a\omega$.
The radial equation has regular singular points at the two zeros
of $\Delta$ and at $\rho=\infty$ ($r=0$), 
and an irregular singular point at $\rho=0$ ($r=\infty$).

%-----------------------------------------------------------------------------
\subsection{Solution to the radial ODE near future null infinity
and near the black hole horizon
\label{sec:radial_solution_near_boundaries}
}
We next show that the ingoing (into the domain) wave solutions 
near future null infinity $(\rho\sim0$) and
near the black hole horizon $(\rho\sim\rho_+)$ are not
smooth at those boundary points.
We also show that smooth solutions to the radial ODE represent 
outgoing/stationary waves at the horizon and future null infinity.
Our notation for the (confluent) hypergeometric function will
follow \cite{beals2016special}; see also \cite{NIST:DLMF} 
for a general reference.

First we consider the behavior of solutions near $\rho=0$
(the irregular singular point of Eq.~\ref{eq:radial_rho}).
To leading order in $\rho$ we have
a confluent hypergeometric equation
\begin{eqnarray}
   -
   \rho^2\frac{d^2R_{\mathcal{J}}}{d\rho^2}
   +
   \left[
      2i\omega
      -
      2\left(1+s\right)\rho
   \right]
   \frac{dR_{\mathcal{J}}}{d\rho}
   \nonumber\\
   +
   \left[
      \left(a^2-16M^2\right)
      \omega^2
      +
      2\left(ma+2isM\right)\omega
      -
      {}_s\Lambda^m_l
   \right]
   R_{\mathcal{J}}
   =
   0
   .
\end{eqnarray}
The general solution to this equation is \cite{beals2016special}
\begin{eqnarray}
   \fl
   R_{\mathcal{J}}\left(\rho\right)
   =
   \left(-\frac{2i\omega}{\rho}\right)^{a_{\mathcal{J}}}
   \Bigg[
      \mathcal{A}_{\mathcal{J}}\times
      M\left(
         a_{\mathcal{J}},
         c_{\mathcal{J}};
         -\frac{2i\omega}{\rho}
      \right)
      +
      \mathcal{B}_{\mathcal{J}}\times
      U\left(
         a_{\mathcal{J}},
         c_{\mathcal{J}};
         -\frac{2i\omega}{\rho}
      \right)
   \Bigg]
   ,
\end{eqnarray}
where $\mathcal{A}_{\mathcal{J}},\mathcal{B}_{\mathcal{J}}$ are constants,
$M,U$ are respectively the confluent hypergeometric functions of the first 
and second kind, and
\begin{eqnarray}
   \fl
   a_{\mathcal{J}}
   \equiv
   \frac{1}{2}\left(1+2s-\sqrt{4c+\left(1+2s\right)^2}\right)
   ,\\
   \fl
   c_{\mathcal{J}}
   \equiv
   1
   -
   \left(
      4\left[
         \left(a^2-16M^2\right)
         \omega^2
         +
         2\left(ma+2isM\right)\omega
         -
         {}_s\Lambda^m_l
      \right]^2
      +
      \left(1+2s\right)^2
   \right)^{1/2}
   .
\end{eqnarray}
In the limit $\rho\to0$, the limiting solution to the function multiplied
by $\mathcal{A}_J$ is (here we have reintroduced the harmonic
time dependence)
\begin{equation}
   \fl 
   \lim_{\rho\to0}
   e^{-i\omega\tau}\left(-\frac{2i\omega}{\rho}\right)^{a_{\mathcal{J}}}
   M\left(
      a_{\mathcal{J}},
      c_{\mathcal{J}};
      -\frac{2i\omega}{\rho}
   \right)
   \sim
   \left(-\frac{2i\omega}{\rho}\right)^{a_{\mathcal{J}}}
   \mathrm{exp}\left(-i\omega\tau-\frac{2i\omega}{\rho}\right)
   \left(\cdots\right)
   .
\end{equation}
This describes a wave solution that is ingoing into the computational domain,
and we see that it is irregular as $\rho\to0$.
To remove the ingoing wave solution we then set $\mathcal{A}_{\mathcal{J}}=0$.
The solution then reads
\begin{equation}
\label{eq:fni_sol}
   R_{\mathcal{J}}\left(\rho\right)
   =
   \mathcal{B}_{\mathcal{J}}\times
   \left(-\frac{2i\omega}{\rho}\right)^{a_{\mathcal{J}}} 
   U\left(
      a_{\mathcal{J}},
      c_{\mathcal{J}};
      -\frac{2i\omega}{\rho}
   \right)
   .
\end{equation}
As the limiting behavior of confluent hypergeometric function of the 
second kind is 
\begin{equation}
   \lim_{\rho\to0}
   U\left(
      a_{\mathcal{J}},
      c_{\mathcal{J}};
      -\frac{2i\omega}{\rho}
   \right)
   \sim
   \left(-\frac{2i\omega}{\rho}\right)^{-a_{\mathcal{J}}}\left(\cdots\right)
   ,
\end{equation}
we see that this solution near future null infinity goes as 
\begin{equation}
   \label{eq:limit_scd_conf_sol}
   \lim_{\rho\to0} e^{-i\omega\tau}R_{\mathcal{J}}\left(\rho\right)
   \sim
   e^{-i\omega\tau}\times\left(
      const.
      +
      \mathcal{O}\left(\rho\right)
   \right)
   .
\end{equation}
From Eq.~\ref{eq:limit_scd_conf_sol}, 
we see that the solution Eq.~\ref{eq:fni_sol} does not
support mode solutions that are ingoing into the computational domain
near $\rho=0$.
At ``worst'' it supports modes that are neither ingoing nor outgoing,
which are consistent with ingoing waves that live exactly at future
null infinity; i.e. modes that have support exactly at $\rho=0$.

We next show that near the black hole horizon $(\rho\sim\rho_+\equiv1/r_+)$,
there are two solutions, one of which is regular and one which is
irregular at $\rho=0$. Furthermore we show that the irregular
solution describes an ingoing wave solution, and the regular
solution describes and outgoing/stationary wave solution.

To leading order 
in $x\equiv\left(1-\rho/\rho_+\right)$, the radial ODE reduces to a 
hypergeometric equation:
\begin{eqnarray}
\label{eq:near_horizon_radial_eqn}
   x\left(\sigma+x\right)
   \frac{d^2R_{\mathcal{H}}}{dx^2}
   +
   \left[
      \frac{\rho_-}{\rho_+^2}A\left(\omega,m,\rho_+\right)
      -
      \left(\frac{\rho_-}{\rho_+}\frac{dA}{d\rho}\Bigg|_{\rho=\rho_+}\right)
      x
   \right]\frac{dR_{\mathcal{H}}}{dx}
   \nonumber\\
   -
   \frac{\rho_-}{\rho_+}\left(
      B\left(\omega,m,\rho_+\right)
      -
      {}_s\Lambda^m_l
   \right)
   R_{\mathcal{H}} 
   =
   0
   ,
\end{eqnarray}
where $\sigma\equiv\left(\rho_-/\rho_+-1\right)$.
Generally we can write the near-horizon solution as
\begin{eqnarray}
   \fl
   R_{\mathcal{H}}(x)
   =
   \mathcal{A}_{\mathcal{H}}\times
   F\left(
      a_{\mathcal{H}},
      b_{\mathcal{H}},
      \frac{c_{\mathcal{H}}}{\sigma};
      -
      \frac{x}{\sigma}
   \right)
   \nonumber\\
   +
   \mathcal{B}_{\mathcal{H}}\times
   \left(\frac{x}{\sigma}\right)^{1-c_{\mathcal{H}}/\sigma}
   F\left(
      1+a_{\mathcal{H}}-\frac{c_{\mathcal{H}}}{\sigma},
      1+b_{\mathcal{H}}-\frac{c_{\mathcal{H}}}{\sigma},
      2-\frac{c_{\mathcal{H}}}{\sigma};
      -
      \frac{x}{\sigma}
   \right)
   ,
\end{eqnarray}
where $\mathcal{A}_{\mathcal{H}},\mathcal{B}_{\mathcal{H}}$ 
are constants, $F$ is the hypergeometric function, and
\begin{eqnarray}
   \label{eq:definitions_near_horizon_1}
   \fl
   c_{\mathcal{H}}
   \equiv
   \frac{\rho_-}{\rho_+^2}
   \Bigg(
      2\left(
         1
         +
         \left(a^2-8M^2\right)\rho_+^2
         +
         4Ma^2\rho_+^3
      \right)i\omega
      \nonumber\\
      \fl
      \qquad\qquad
      -
      2\left(1+s\right)\rho_+
      +
      2\left[
         ima
         +
         \left(s+3\right)M
      \right]\rho_+^2
      -
      4a^2\rho_+^3
   \Bigg)
   \\
   \label{eq:definitions_near_horizon_2}
   \fl
   1+a_{\mathcal{H}}+b_{\mathcal{H}}
   \equiv
   \frac{\rho_-}{\rho_+}\Bigg(
      \left(
         -
         24a^2M\rho_+^2
         -
         4a^2\rho_+
         +
         32M^2\rho_+
      \right)i\omega
      \nonumber\\
      \fl
      \qquad\qquad\qquad\qquad
      +
      2\left(1+s\right)
      +
      12a^2\rho_+^2
      -
      4ima\rho_+
      -
      4\left(3+s\right)M\rho_+
   \Bigg)
   ,\\
   \label{eq:definitions_near_horizon_3}
   \fl
   a_{\mathcal{H}}b_{\mathcal{H}}
   \equiv
      -
      \left(
         \left(a^2-16M^2\right)\omega^2
         +
         2\left(ma+2isM\right)\omega
      \right)\frac{\rho_-}{\rho_+}
      \nonumber\\
      \fl
      -
      2\left[
         4\left(a^2-4M^2\right)M\omega^2
         +
         4\left(m aM - i\left(s+2\right)M^2 + ia^2\right)\omega
         +
         ima
         +
         \left(s+1\right)M
      \right]
      \rho_-
      \nonumber\\
      \fl
      -
      2\left(
         8a^2M^2\omega^2
         +
         6ia^2M\omega
         -
         a^2
      \right)
      \rho_+\rho_-
   .
   ,
\end{eqnarray}
Near $x\sim0$, the term multiplied by $\mathcal{B}_{\mathcal{H}}$ 
goes as (here we have reintroduced the harmonic time dependence) 
\begin{equation}
   \label{eq:b_term_ingoing}
   \sim \mathcal{B}_{\mathcal{H}}\times \mathrm{exp}\left[
      -
      i\omega \tau 
      + 
      \left(\frac{c_{\mathcal{H}}}{\sigma}-1\right)\log\frac{\sigma}{x}
   \right]
   \left(\cdots\right)
   .
\end{equation}
We see that this solution describes waves that oscillate rapidly near
$x=0$, and are ingoing into the computational domain provided the
component $c_{\mathcal{H}}/\sigma - 1$ which multiplies $i\omega$
is negative. Examining
\begin{equation}
   \frac{c_{\mathcal{H}}}{\sigma}-1
   =
   \left(
      \frac{2}{\sigma}\frac{\rho_-}{\rho_+^2}\left(
         1
         +
         \left(a^2-8M^2\right)\rho_+^2
         +
         4Ma^2\rho^3_+
      \right)
   \right)
   i\omega
   +
   \cdots
   ,
\end{equation}
and noting that $\rho_-,\rho_+,\sigma\geq0$, 
we see that we only need to determine
if $
   \left(
      1
      +
      \left(a^2-8M^2\right)\rho_+^2
      +
      4Ma^2\rho^3_+
   \right)
   $
is negative.
Using $\rho_+\equiv1/r_+=1/(M+\sqrt{M^2-a^2})$,
it is straightforward to verify that this is true
for all $a\in[0,M]$.
Thus \ref{eq:b_term_ingoing} describes an ingoing mode solution. 
The regular solution limits to a constant as $x\to0$: 
\begin{equation}
   \label{eq:near_horizon_solution}
   \mathcal{R}_{\mathcal{H}}
   =
   \mathcal{A}_{\mathcal{H}}\times
   F\left(
      a_{\mathcal{H}},
      b_{\mathcal{H}},
      \frac{c_{\mathcal{H}}}{\sigma};
      -
      \frac{x}{\sigma}
   \right)
   =
   \mathcal{A}_{\mathcal{H}}
   +
   \mathcal{O}\left(x\right)
   .
\end{equation}
Similar to the regular solution at $\rho=0$ (Eq.~\ref{eq:fni_sol}),
we see that the solution Eq.~\ref{eq:near_horizon_solution} does not
support mode solutions that are ingoing into the computational domain.
At ``worst'' it supports modes that are neither ingoing nor outgoing,
which are consistent with outgoing waves that live exactly at the black
hole horizon; i.e. support exactly at $\rho=\rho_+$.

To conclude, we have shown that the radial Teukolsky equation admits 
a regular and an irregular solution at the two boundary points $\rho=0,\rho_+$.
We have shown that the irregular solutions represent 
modes which are ingoing into the computational domain,
while the regular solutions do not support such modes.
Thus the physical boundary conditions for the radial equation
are to impose regularity at $\rho=0,\rho_+$.
%%%%%%%%%%%%%%%%%%%%%%%%%%%%%%%%%%%%%%%%%%%%%%%%%%%%%%%%%%%%%%%%%%%%%%%%%%%%%%
\section{Discretization of the angular and radial ODEs, and numerically
   computing the quasinormal modes
   \label{sec:discretization_method}
}
We discretize the radial equation, Eq.~\ref{eq:radial_rho}, 
using Chebyshev pseudospectral (collocation) methods 
(e.g. \cite{trefethen2000spectral,boyd2001chebyshev}).
These methods automatically impose regularity of the solution at 
the boundaries of the domain $\rho=0,\rho_+$\footnote{We
note that this method has been used to compute 
the QNMs and QNEs for spherically symmetric black hole spacetimes
\cite{Jansen:2017oag}.}.
We briefly review Chebyshev pseudospectral methods in
\ref{sec:radial_ODE_pseudospectral}.

We discretize the angular equation, Eq.~\ref{eq:spheroidal_laplacian},
using a spectral method \cite{Cook:2014cta} 
(see also Appendix A of \cite{Hughes:1999bq}).
We expand the spin-weighted spheroidal harmonics as a linear sum
of spin-weighted spherical harmonics, and evaluate
the angular ODE in coefficient/spectral space. In spectral space
the angular ODE reduces to a sparse, banded matrix equation.
For completeness, we briefly review this method in \ref{sec:angular_ODE_sparse}.

Fixing the labels $(s,l,m)$,
the discretized radial and angular ODEs respectively take the form 
\begin{eqnarray}
   \label{eq:discretized_radial}
   \sum_{j=0}^{N_{(\rho)}}\left(
      \left[\hat{M}_{(\rho)}\left(\omega\right)\right]_{ij}
      -
      \Lambda_{(\rho)}\hat{I}_{ij}
   \right)
   \vec{f}_j
   =
   0
   ,\\
   \label{eq:discretized_angular}
   \sum_{j=0}^{N_{(\theta)}}\left(
      \left[\hat{M}_{(\theta)}\left(\omega\right)\right]_{ij}
      -
      \Lambda_{(\theta)}\hat{I}_{ij}
   \right)
   \vec{g}_j
   =
   0
   .
\end{eqnarray}
The matrices $\hat{M}_{(\rho)}$ and $\hat{M}_{(\theta)}$
are functions of powers of $\rho,\omega$ as can be seen
respectively from
Eq.~\ref{eq:radial_rho} and Eq.~\ref{eq:spheroidal_laplacian},
and $\Lambda_{(\rho/\theta)}$ are 
the separation constants ${}_s\Lambda^m_l\left(a\omega\right)$.
We view the system 
Eq.~\ref{eq:discretized_radial}, Eq.~\ref{eq:discretized_angular}
as an eigenvalue equation for the angular separation constant:
the QNMs are the $\omega$ such that
$\Lambda_{(\rho)}=\Lambda_{(\theta)}$, and the QNE 
are the eigenvectors corresponding to eigenvalues $\Lambda$.
To compute the QNMs and QNEs then, we first search
for the zeros to the function
\begin{equation}
   F(\omega)
   \equiv
   \left|
      \Lambda_{(\rho)}\left(\omega\right)
      -
      \Lambda_{(\theta)}\left(\omega\right)
   \right|
   .
\end{equation}
We choose $\Lambda_{(\rho)}$ to be the smallest (in absolute magnitude)
eigenvalue from the radial system, and choose $\Lambda_{(\theta)}$
to be the $l^{th}$ smallest (in absolute value) eigenvalue from the
angular equation.
We search for the zeros of $F$ using Newton's method
\begin{equation}
   \omega_{(n+1)}
   =
   \omega_{(n)}
   -
   \gamma\frac{F}{F'}\Big|_{\omega=\omega_{(n)}}
   ,
\end{equation}
where $0<\gamma\leq1$, and $F'$ denotes
the complex derivative of $F$, which we compute using a second-order
accurate finite difference stencil (here $0<\epsilon\ll1$ is a real number):
\begin{equation}
   \label{eq:fd_formula}
   \fl
   F'\big|_{\omega=\omega_{(n)}}
   \approx
   \frac{
      F\left(\omega_{(n)} + \epsilon\right)
      -
      F\left(\omega_{(n)} - \epsilon\right)
   }{
      2\epsilon
   }
   +
   \frac{
      F\left(\omega_{(n)} + i\epsilon\right)
      +
      F\left(\omega_{(n)} - i\epsilon\right)
   }{
      2i\epsilon
   }
.
\end{equation}
In our search for QNMs, we can exclude modes with positive
imaginary part as there are no such mode solutions 
for the Teukolsky equation, even in the extremal limit 
\cite{Whiting:1988vc,TeixeiradaCosta:2019skg}.

The angular separation constants $\Lambda$
and QNMs $\omega$
have an unambiguous labeling in azimuthal angular number
$m$ and black hole spin $a/M$,
but there is some ambiguity in their labeling $(n,l)$; for more discussion
see \cite{Cook:2014cta}. 
%%%%%%%%%%%%%%%%%%%%%%%%%%%%%%%%%%%%%%%%%%%%%%%%%%%%%%%%%%%%%%%%%%%%%%%%%%%%%%
\section{Results: quasinormal modes and quasinormal eigenfunctions 
   \label{sec:quasinormal_mode_eigenfunctions}
}
   We next present several example QNEs which we
computed using a code that implements the algorithm described in 
Sec.~\ref{sec:discretization_method} \cite{code_online}. 
In the plots we show, we began with ``seed'' values for $(\omega,\Lambda)$
%for the $(s,n,l,m)$ QNM by using the value from
computed from
the \texttt{qnm} package for that mode, and then let the root-finding
algorithm in our code relax to the final $(\omega,\Lambda)$. 
We find the resulting $(\omega,\Lambda)$ to
be identical (to within numerical precision)
to the value given by the \texttt{qnm} code (see Table.~\ref{table:qnm}).
We normalize the radial eigenfunctions so that their maximum amplitude
is equal to one.

%-----------------------------------------------------------------------------
\subsection{Moderate black hole spin quasinormal mode eigenfunctions}

\begin{figure*}[h]
\begin{center}
    \includegraphics[width=0.45\columnwidth]{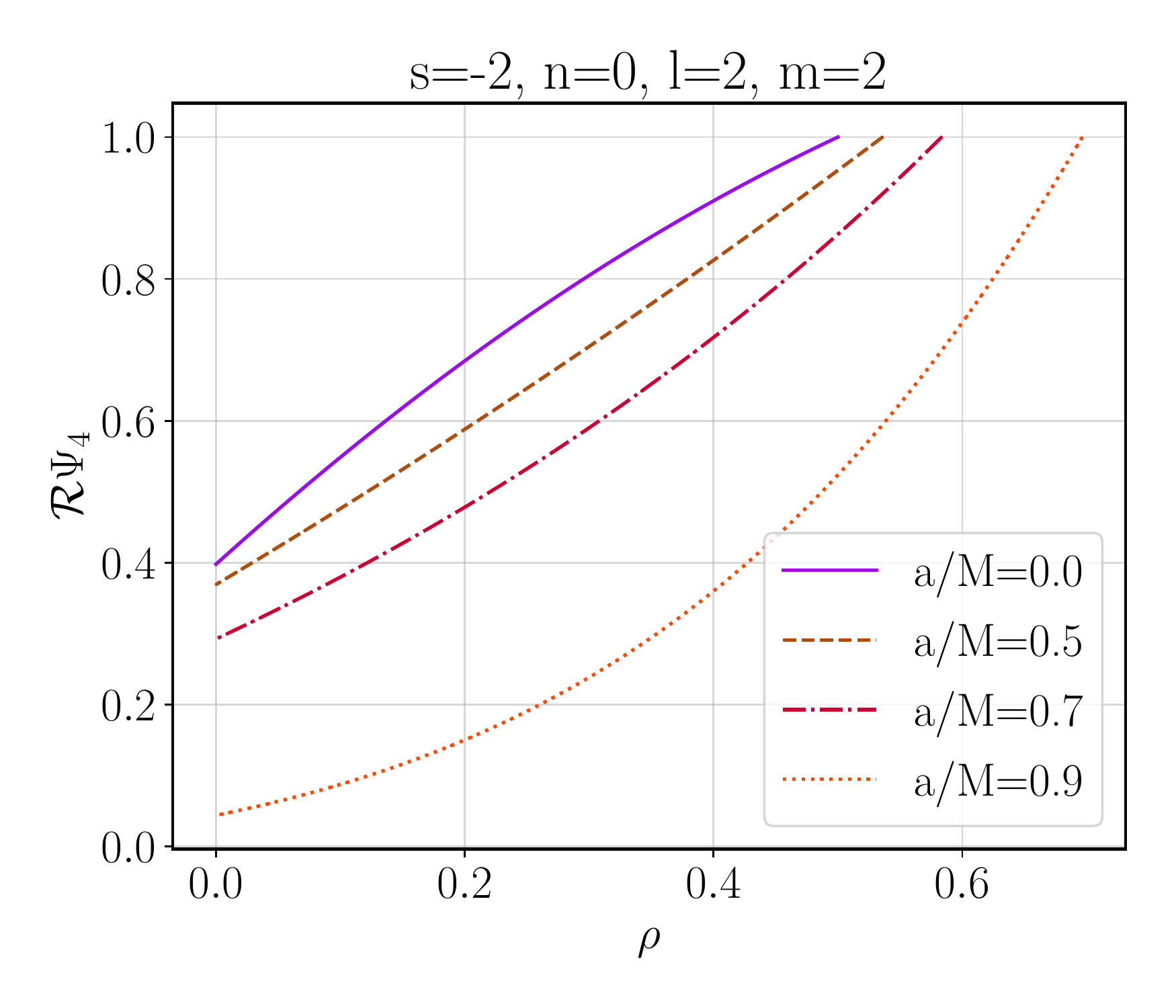}
    \includegraphics[width=0.45\columnwidth]{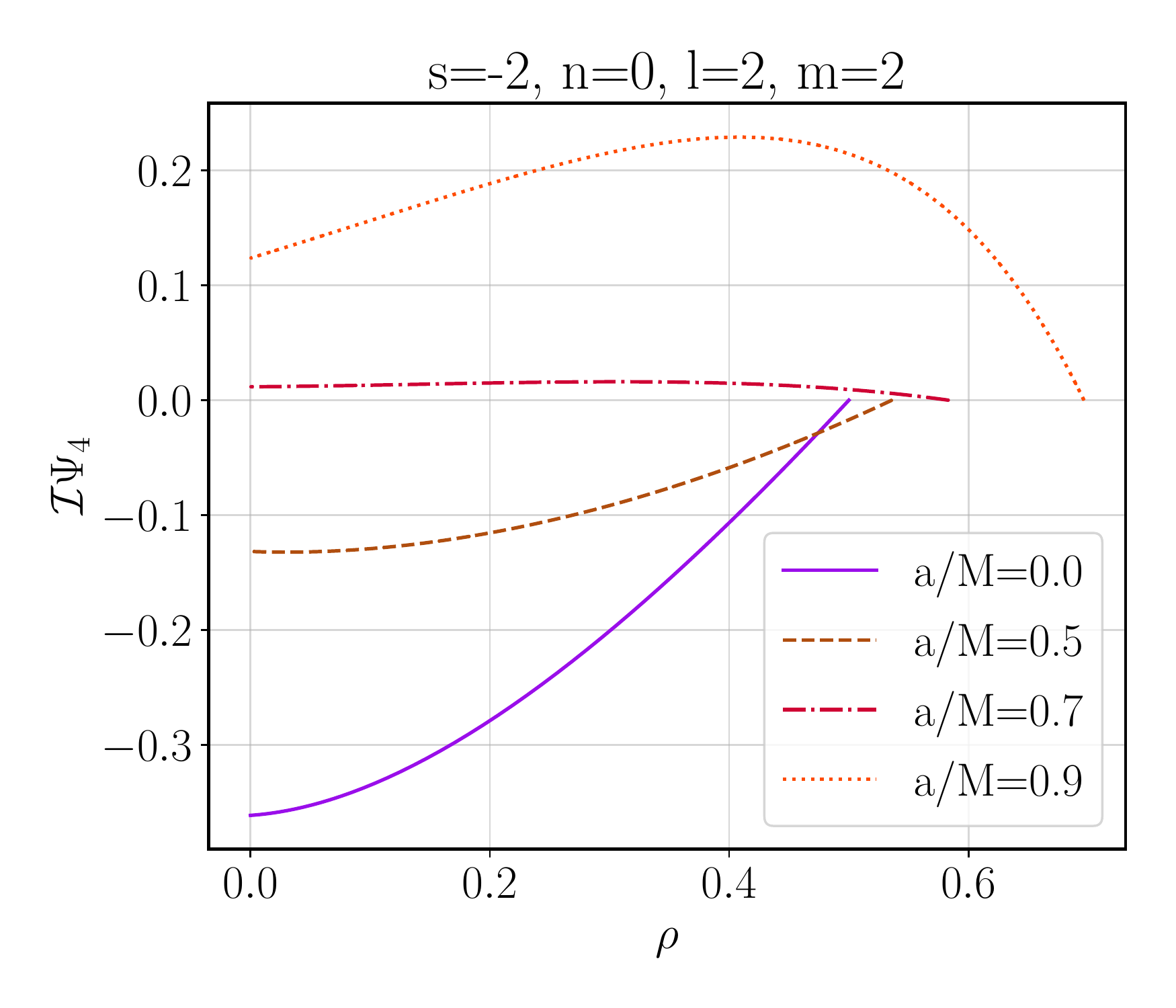}
    \includegraphics[width=0.45\columnwidth]{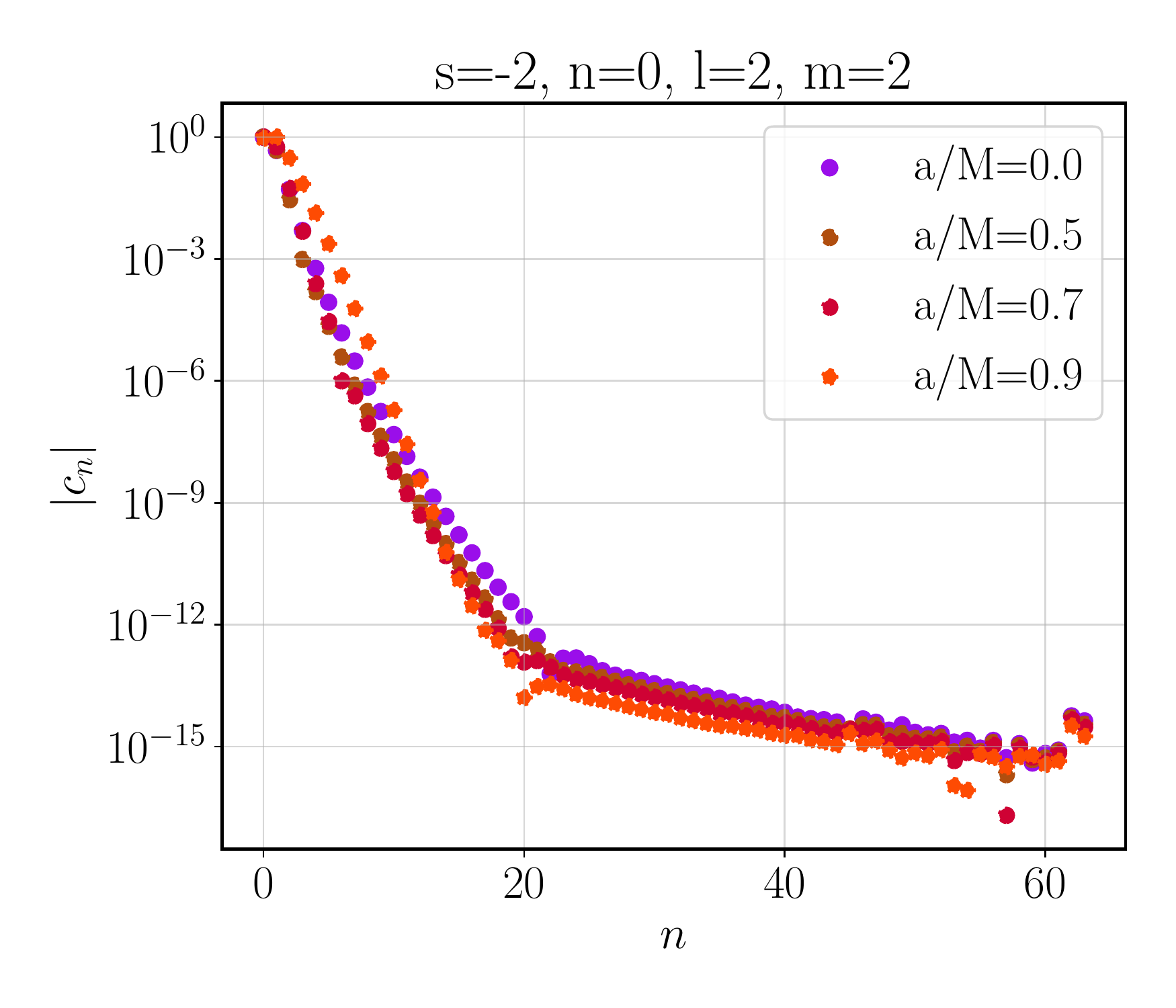}
    \includegraphics[width=0.45\columnwidth]{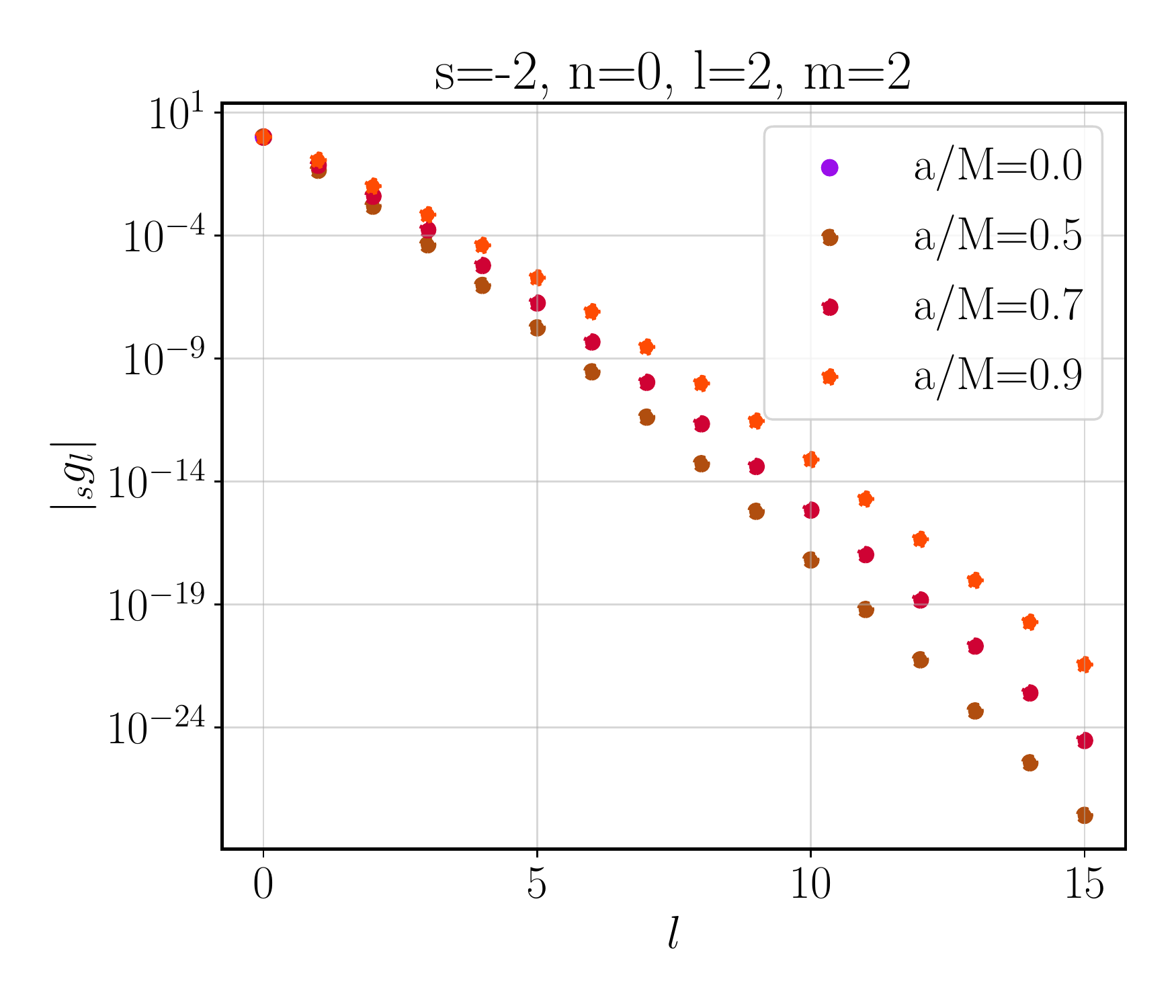}
\end{center}
   \caption{
      The $s=-2$ QNEs for relatively low spins.
      Future null infinity is located at $\rho=0$, and the black hole
      horizon is located at $\rho=\rho_+$, which changes with the 
      black hole spin.
      The lower left panel plots the absolute value of the Chebyshev coefficients,
      $|c_n|$, used to fit the radial mode, and the lower right panel
      shows the absolute value of the spin-weighted spherical harmonic
      coefficients for the QNE.
   }
\label{fig:sm2_lowspin}
\end{figure*}

\begin{figure*}[h]
\begin{center}
    \includegraphics[width=0.45\columnwidth]{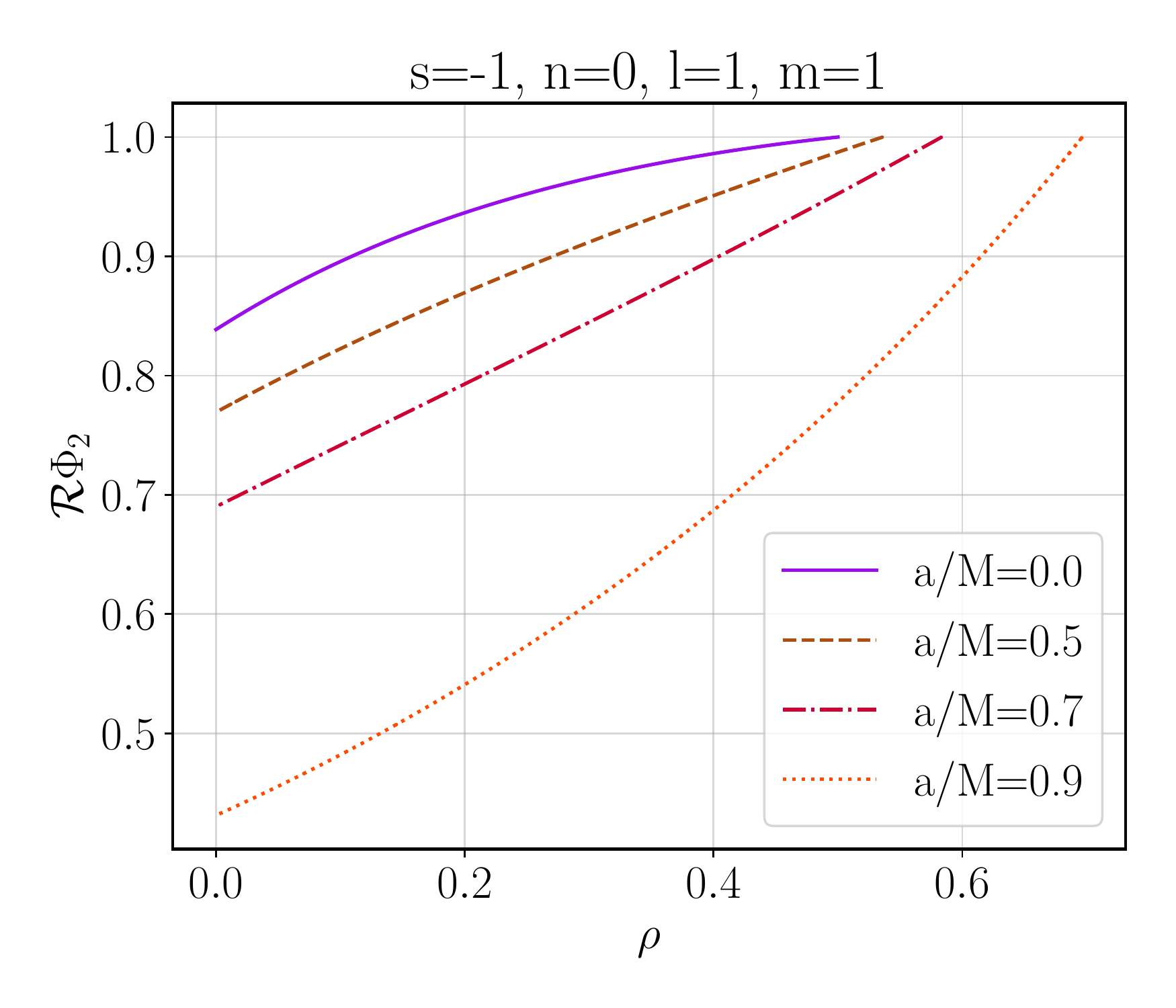}
    \includegraphics[width=0.45\columnwidth]{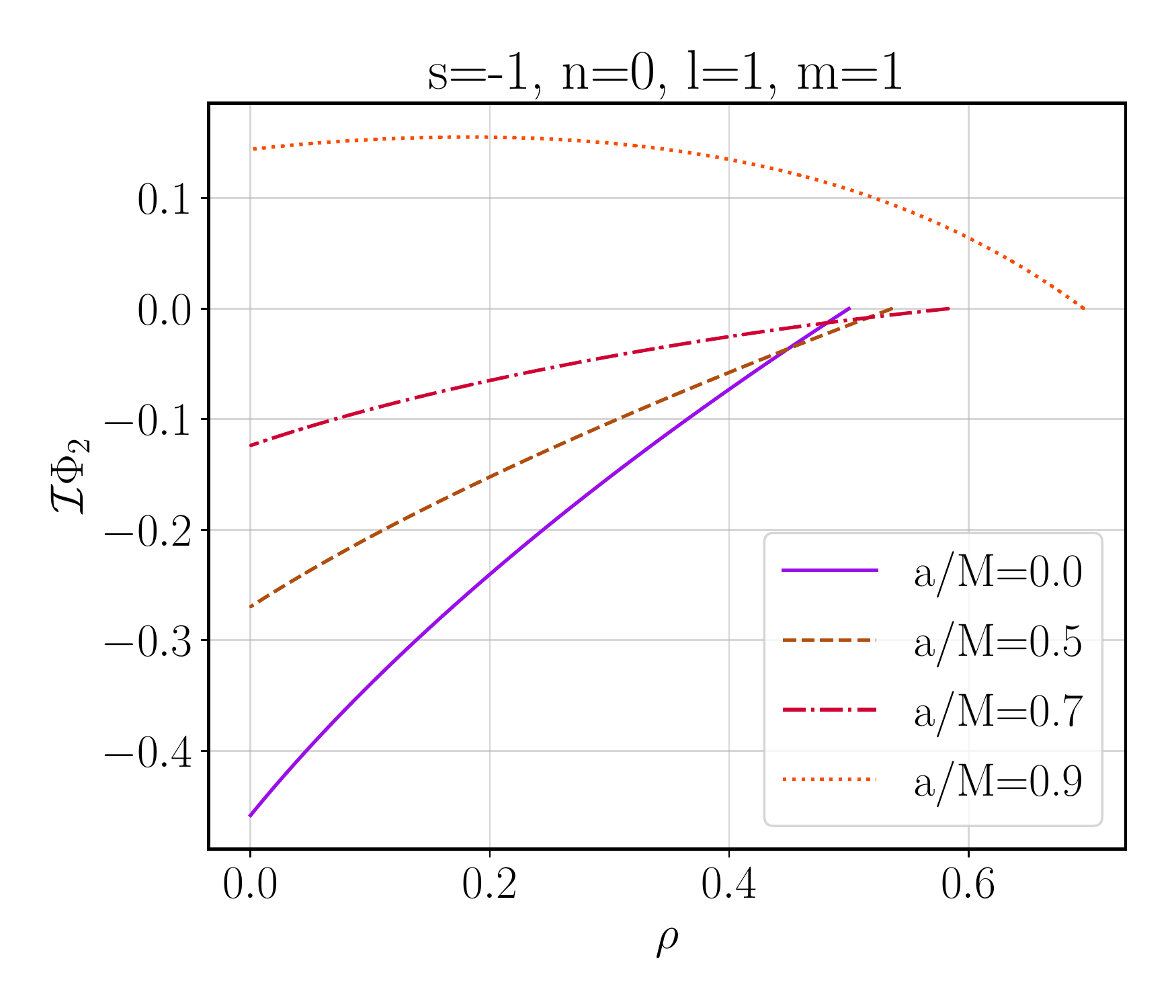}
    \includegraphics[width=0.45\columnwidth]{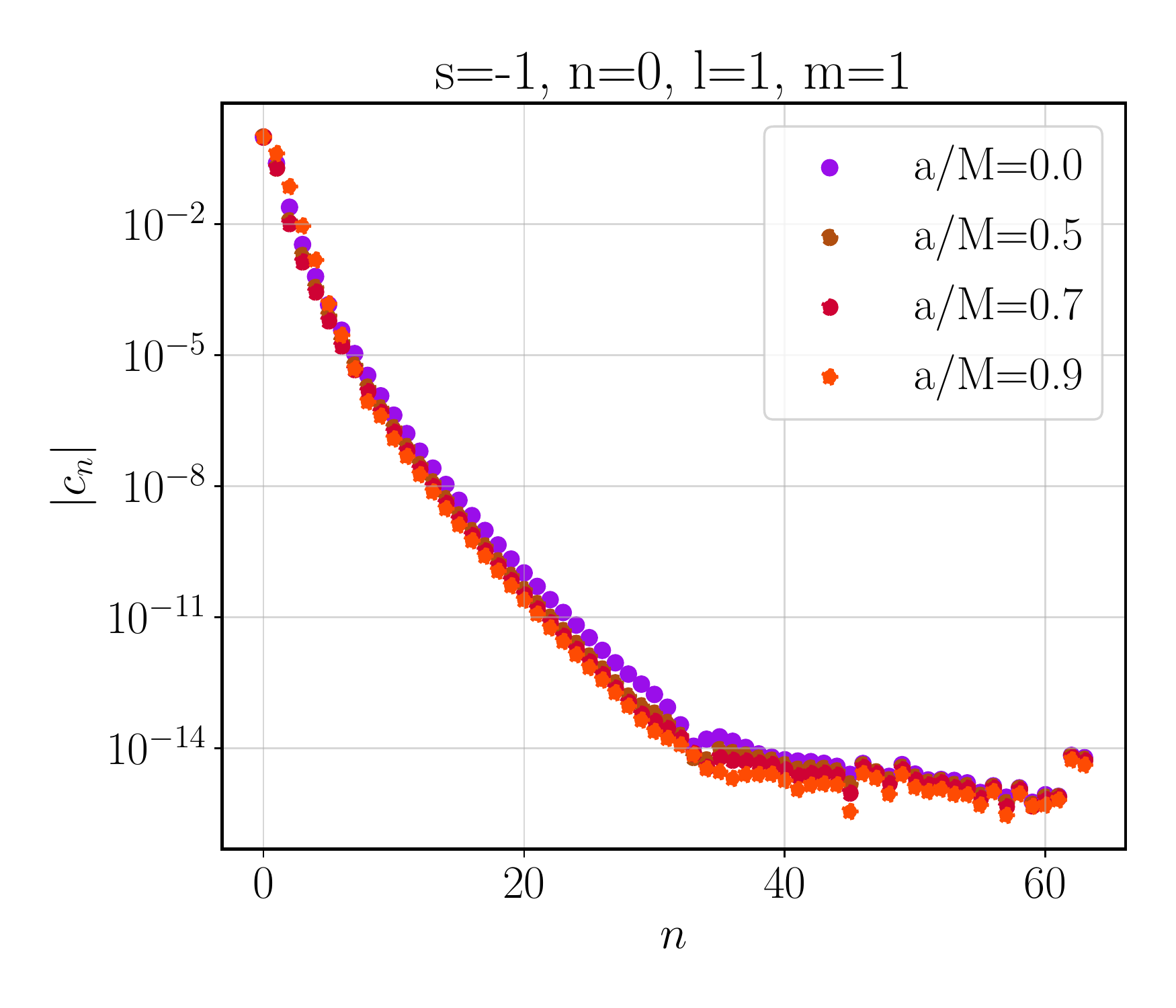}
    \includegraphics[width=0.45\columnwidth]{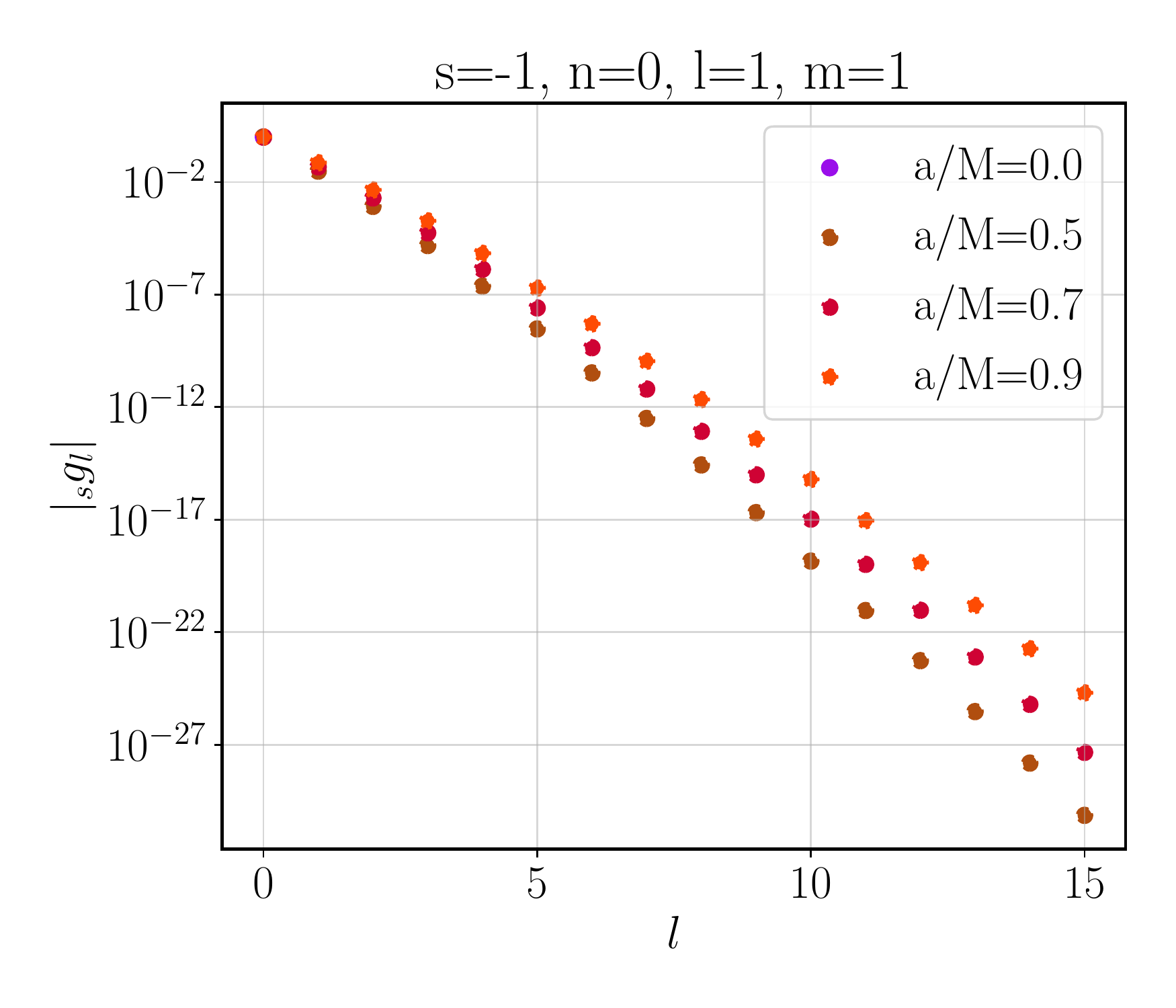}
\end{center}
   \caption{
      The $s=-1$ QNEs for relatively low spins.
      Future null infinity is located at $\rho=0$, and the black hole
      horizon is located at $\rho=\rho_+$, which changes with the 
      black hole spin.
      The lower left panel plots the absolute value of the Chebyshev coefficients,
      $|c_n|$, used to fit the radial mode, and the lower right panel
      shows the absolute value of the spin-weighted spherical harmonic
      coefficients for the QNE.
   }
\label{fig:sm1_lowspin}
\end{figure*}

In Fig.~\ref{fig:sm2_lowspin} we plot the QNEs 
for the $s=-2,n=0,l=2,m=2$ QNMs,
for small to moderately large black hole spins.
In the upper two panels we present the real
and imaginary parts of the radial part of the QNE.
In the lower left panel we plot the absolute value of the
Chebyshev coefficients for the radial part of the QNE. 
Finally, in the lower right panel we plot the absolute value of the
spin-weighted spherical harmonic coefficients for the QNE. 
For slowly/moderately spinning black holes, 
we see that with around
$20$ Chebyshev and spin-weighted spherical harmonics we can describe
the eigenfunctions for the fundamental $s=-2,l=2,m=2$ mode to high precision.
We show a similar set of figures for the $s=-1,n=0,l=1,m=1$ quasinormal modes
in Fig.~\ref{fig:sm1_lowspin}.


%-----------------------------------------------------------------------------
\subsection{Near-extremal quasinormal mode solutions}

\begin{figure*}[h]
\begin{center}
    \includegraphics[width=0.45\columnwidth]{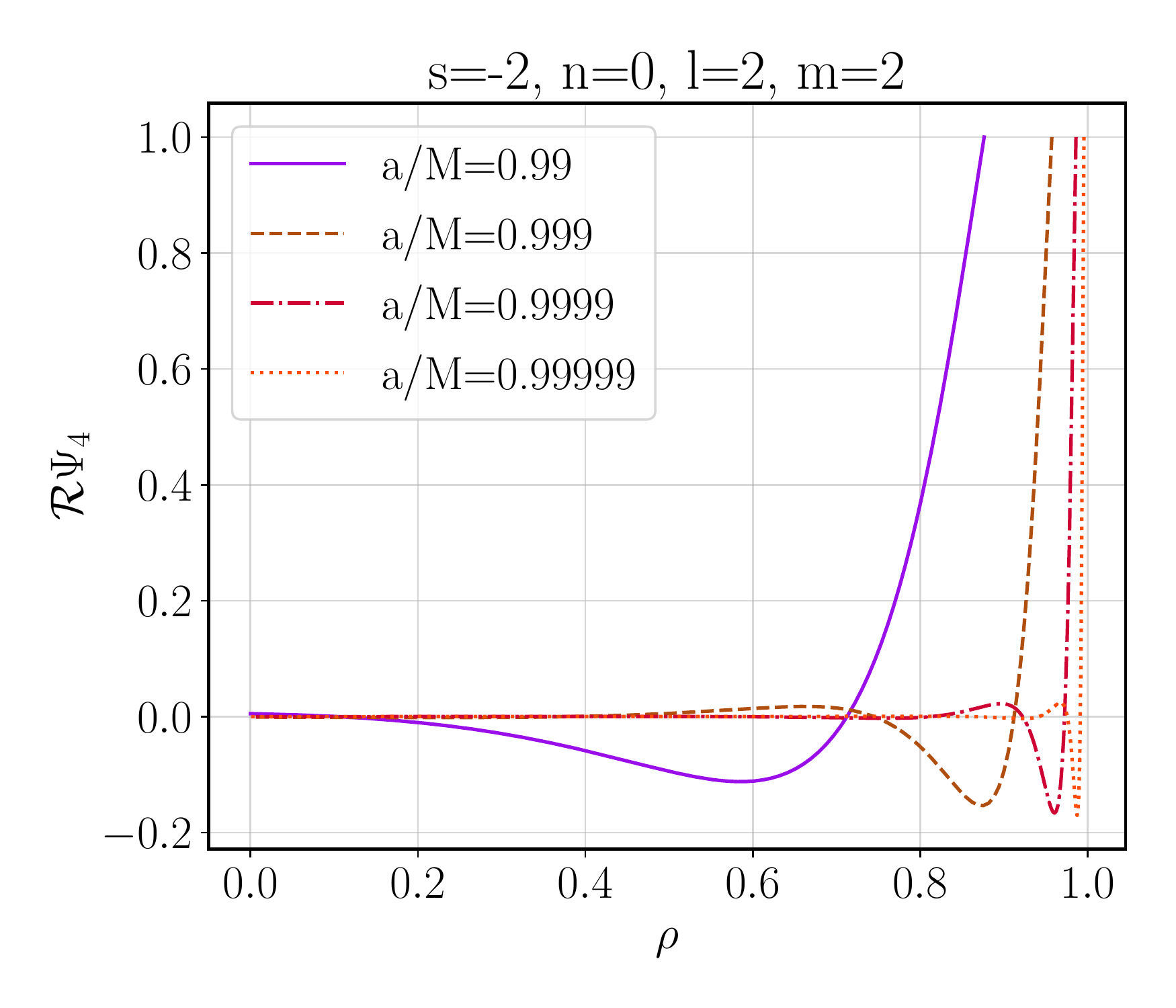}
    \includegraphics[width=0.45\columnwidth]{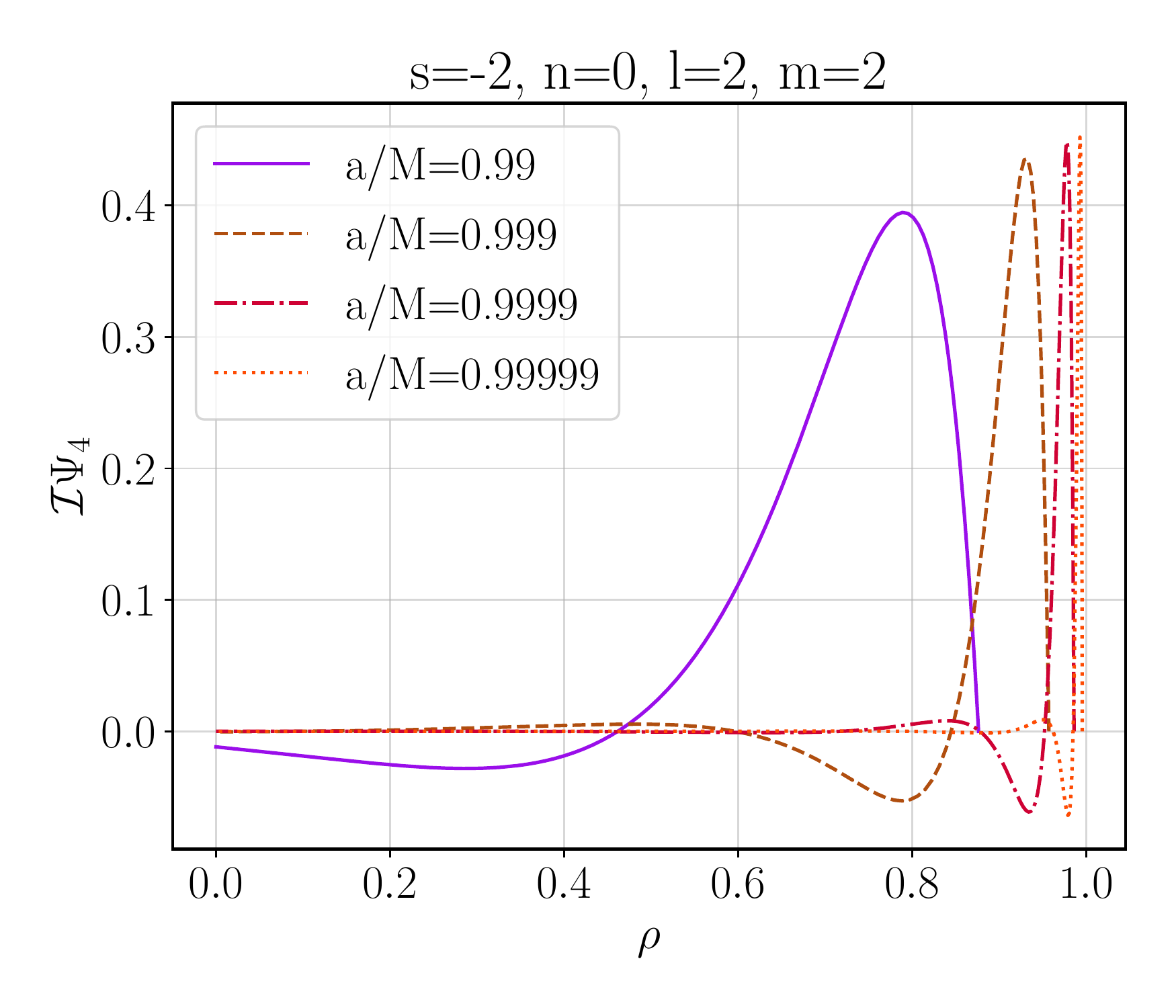}
    \includegraphics[width=0.45\columnwidth]{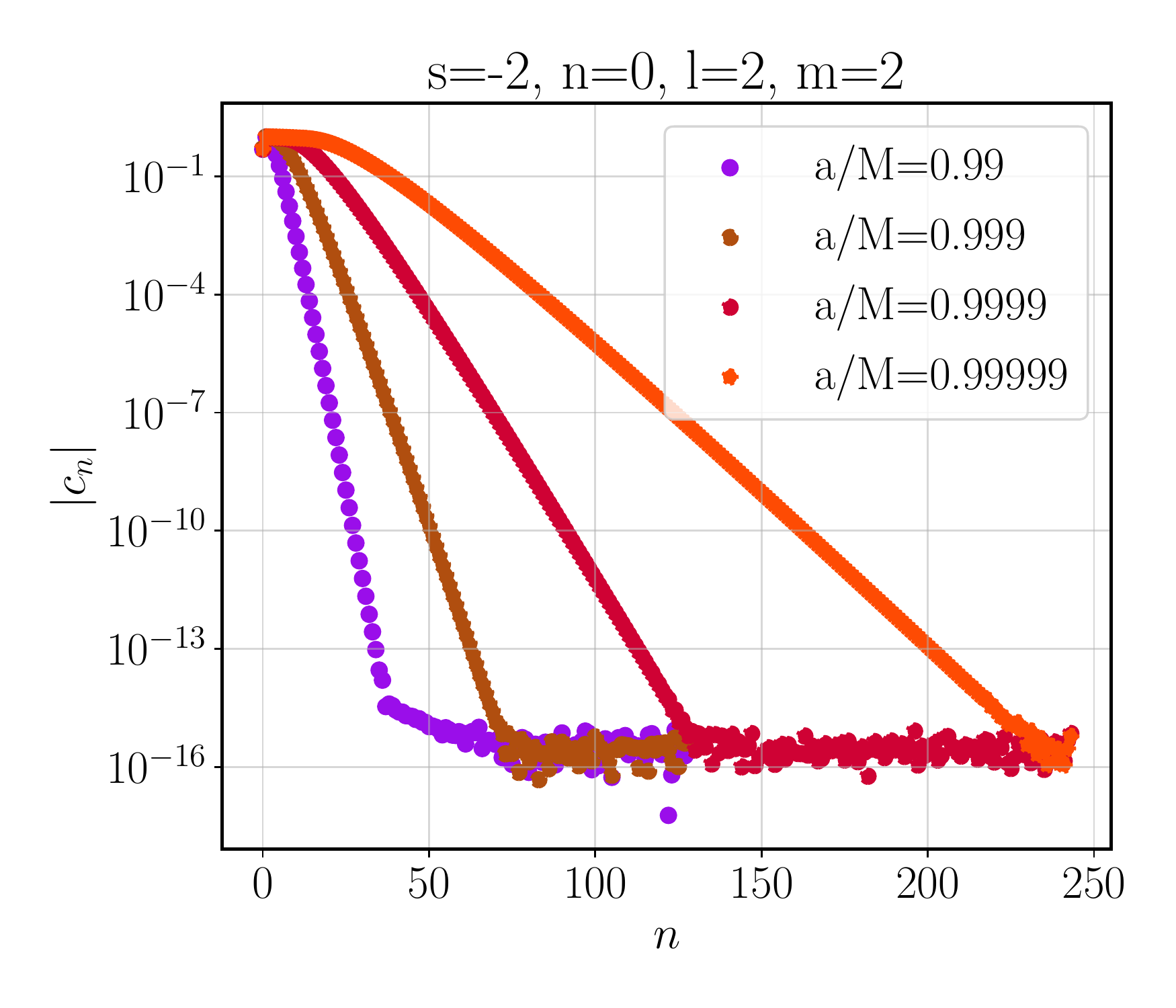}
    \includegraphics[width=0.45\columnwidth]{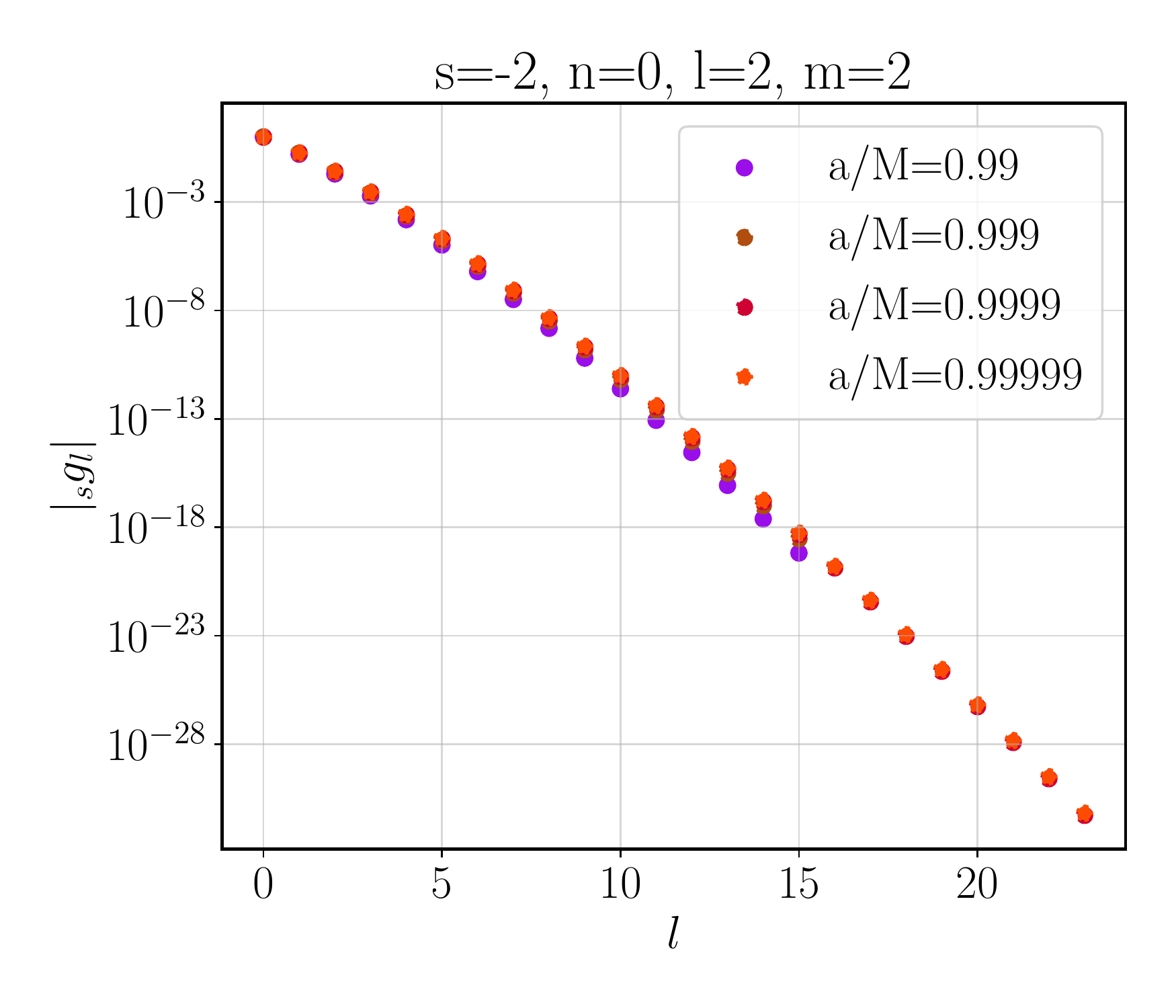}
\end{center}
   \caption{
      The $s=-2$ QNEs 
      in the limit of relatively high black hole spins.
      Future null infinity is located at $\rho=0$, and the black hole
      horizon is located at $\rho=\rho_+$, which changes with the 
      black hole spin.
      As we increase the black hole spin, we need to increase the resolution
      in the radial direction, but not significantly in the
      angular direction.
      We see that as $a\to 1$, the QNEs become
      localized near the black hole horizon.
   }
\label{fig:sm2_highspin}
\end{figure*}

\begin{figure*}[h]
\begin{center}
    \includegraphics[width=0.45\columnwidth]{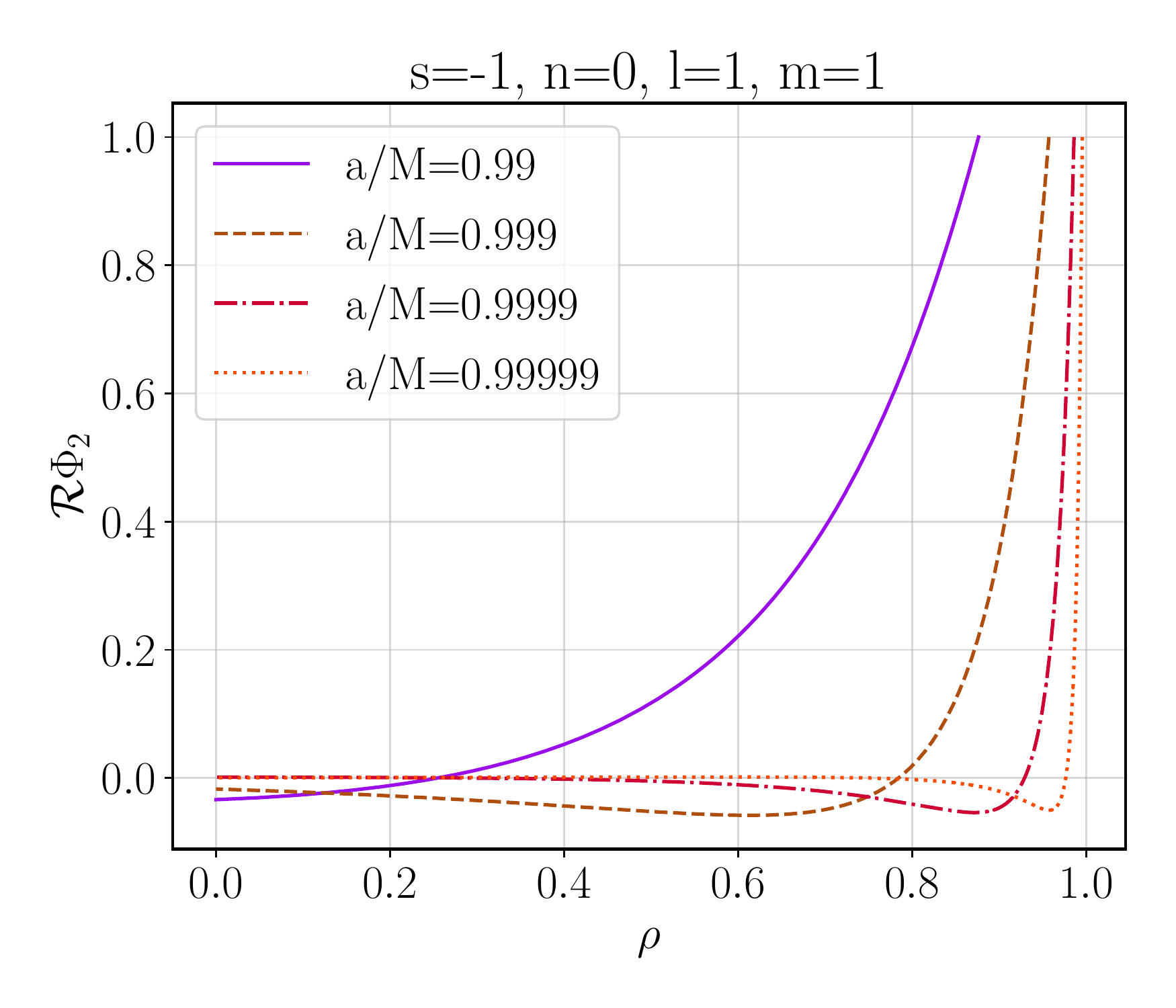}
    \includegraphics[width=0.45\columnwidth]{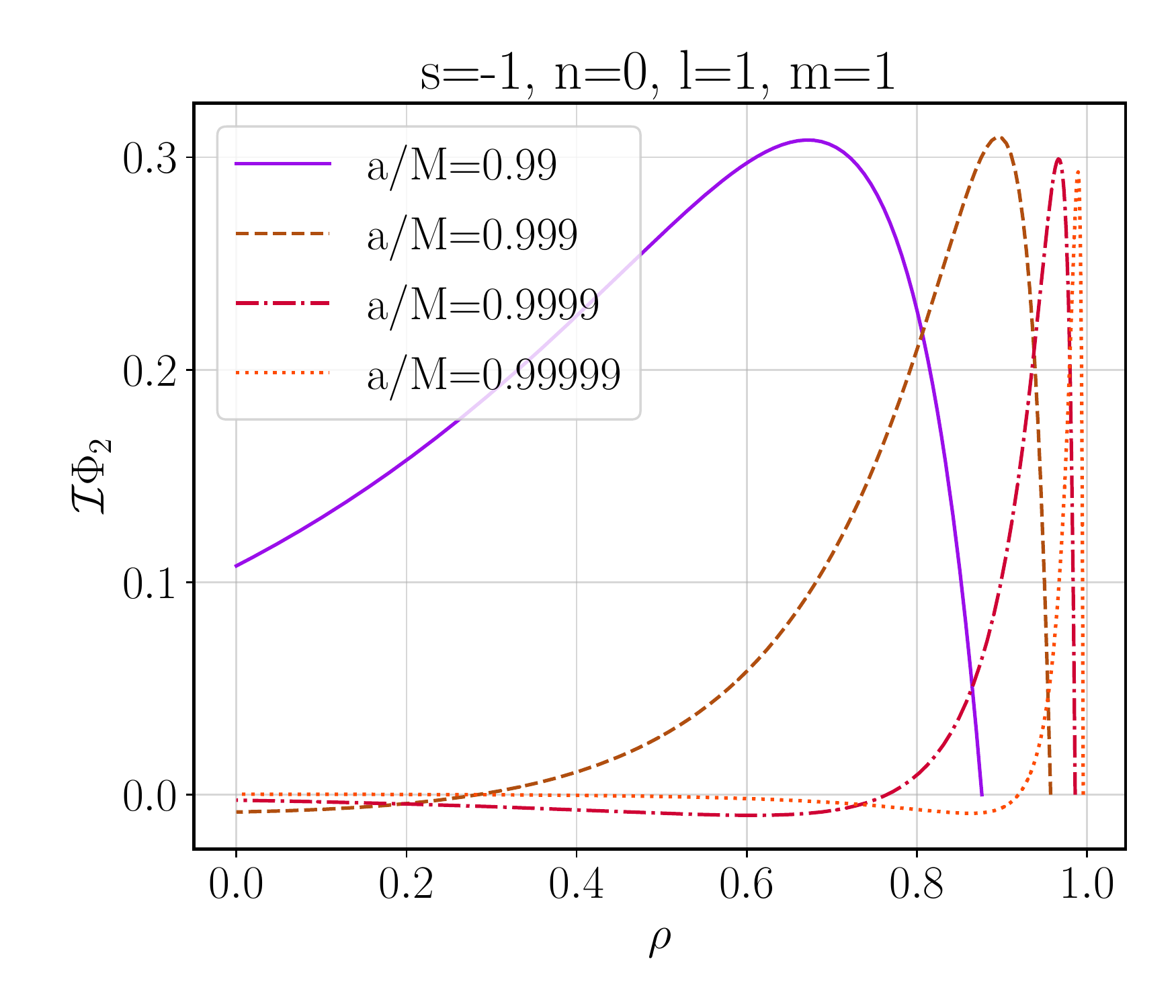}
    \includegraphics[width=0.45\columnwidth]{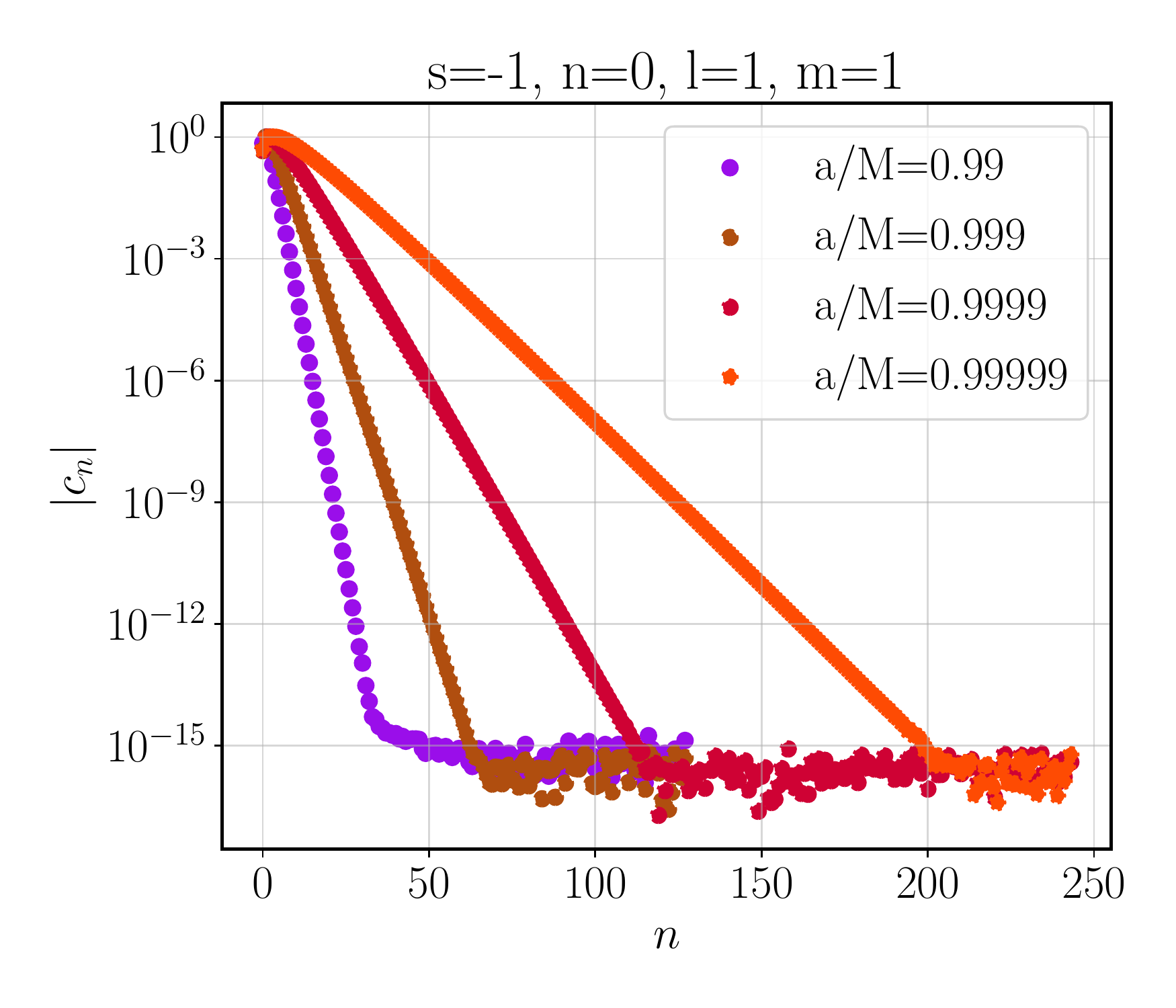}
    \includegraphics[width=0.45\columnwidth]{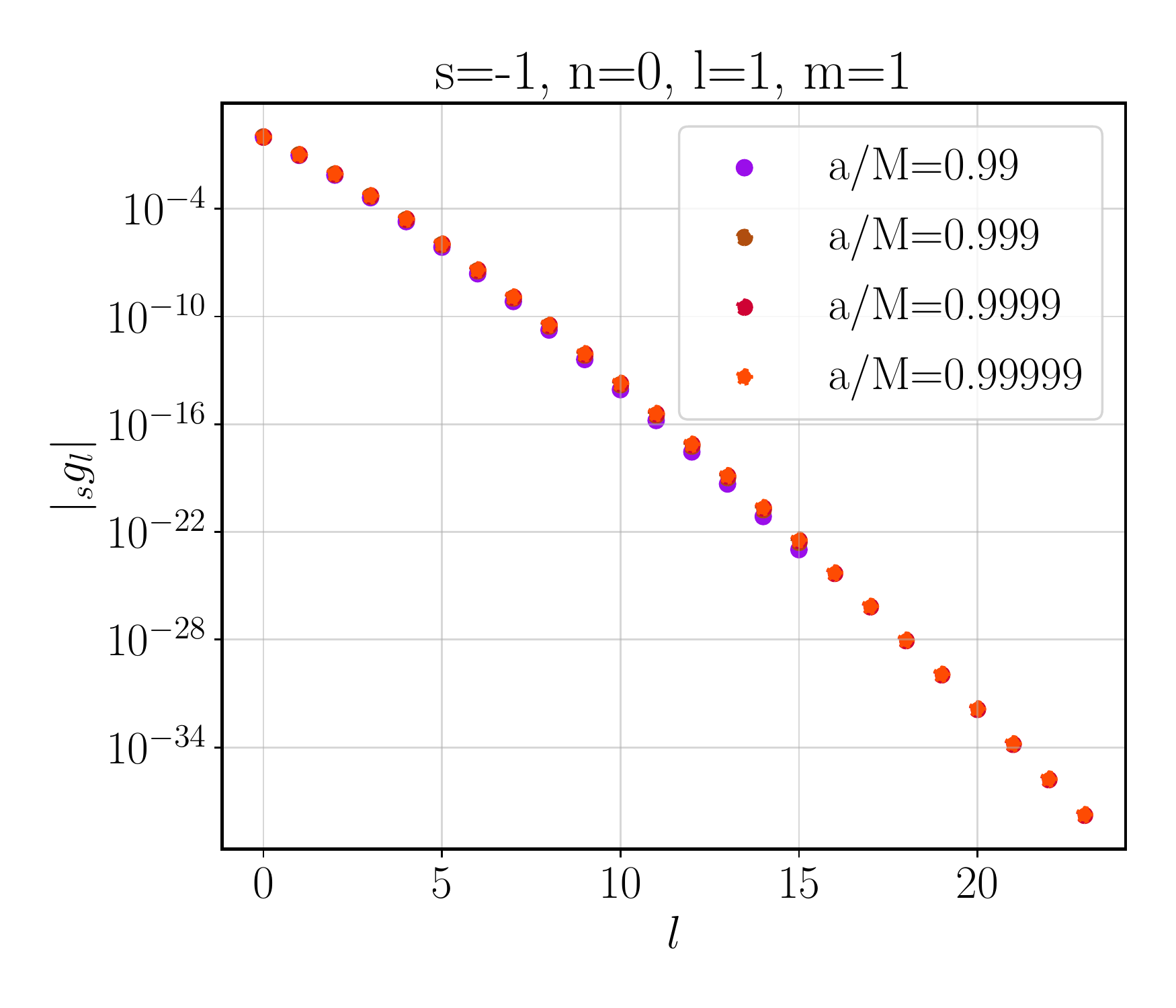}
\end{center}
   \caption{
      The $s=-1$ QNEs 
      in the limit of relatively high black hole spins.
      Future null infinity is located at $\rho=0$, and the black hole
      horizon is located at $\rho=\rho_+$, which changes with the 
      black hole spin.
      As we increase the black hole spin, we need to increase the resolution
      in the radial direction, but not significantly in the
      angular direction.
      We see that as $a\to 1$, the QNEs become
      localized near the black hole horizon.
   }
\label{fig:sm1_highspin}
\end{figure*}

\begin{figure*}[h]
\begin{center}
    \includegraphics[width=0.45\columnwidth]{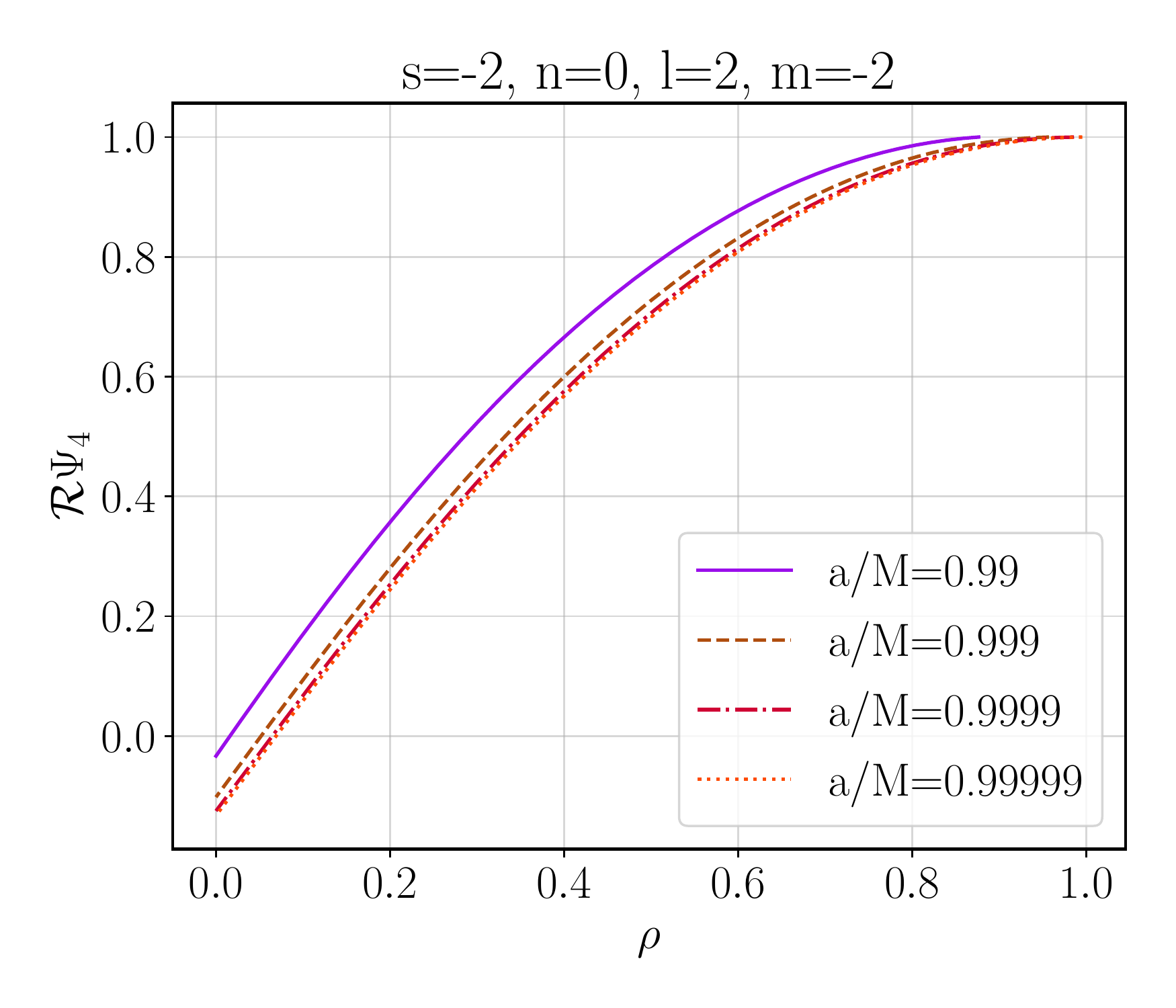}
    \includegraphics[width=0.45\columnwidth]{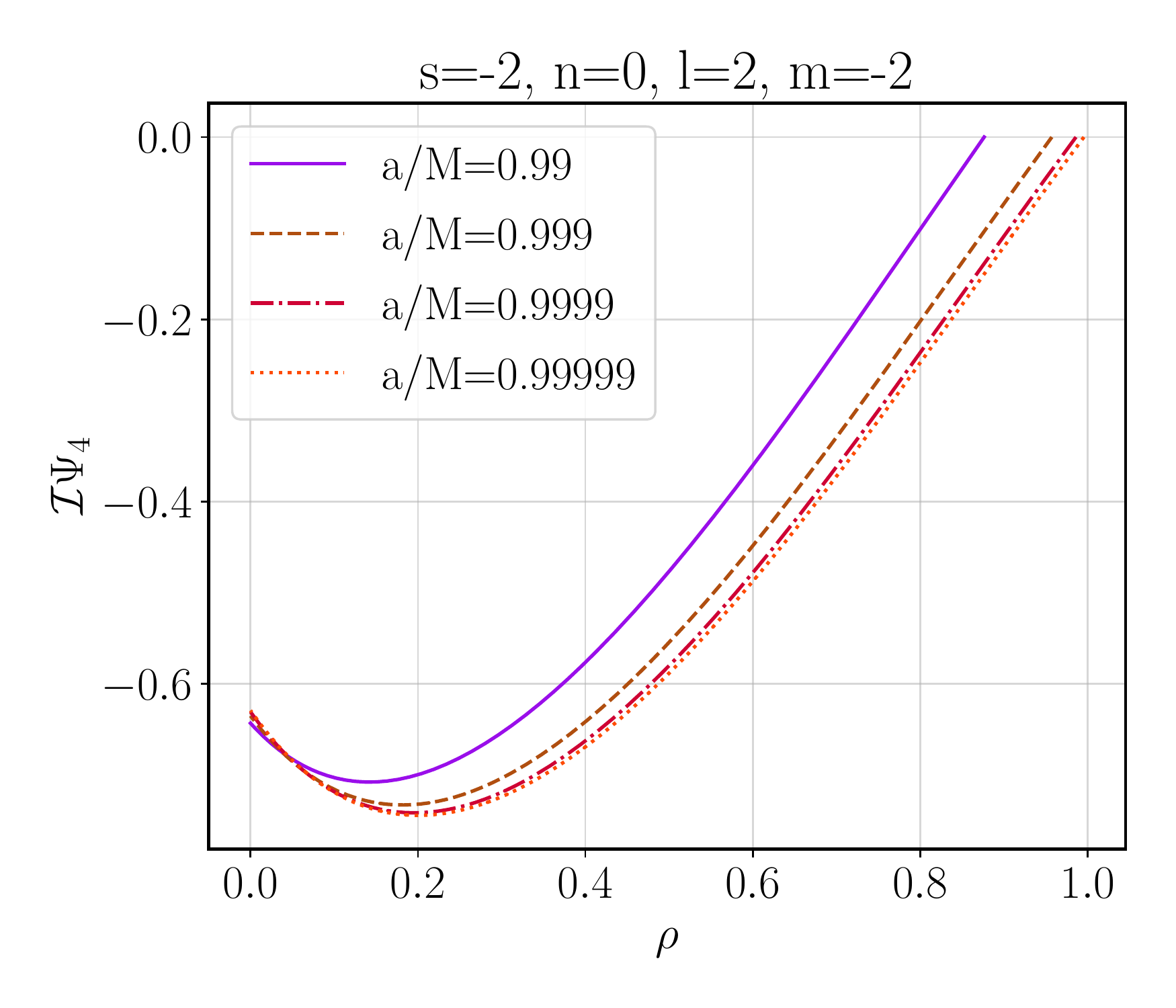}
    \includegraphics[width=0.45\columnwidth]{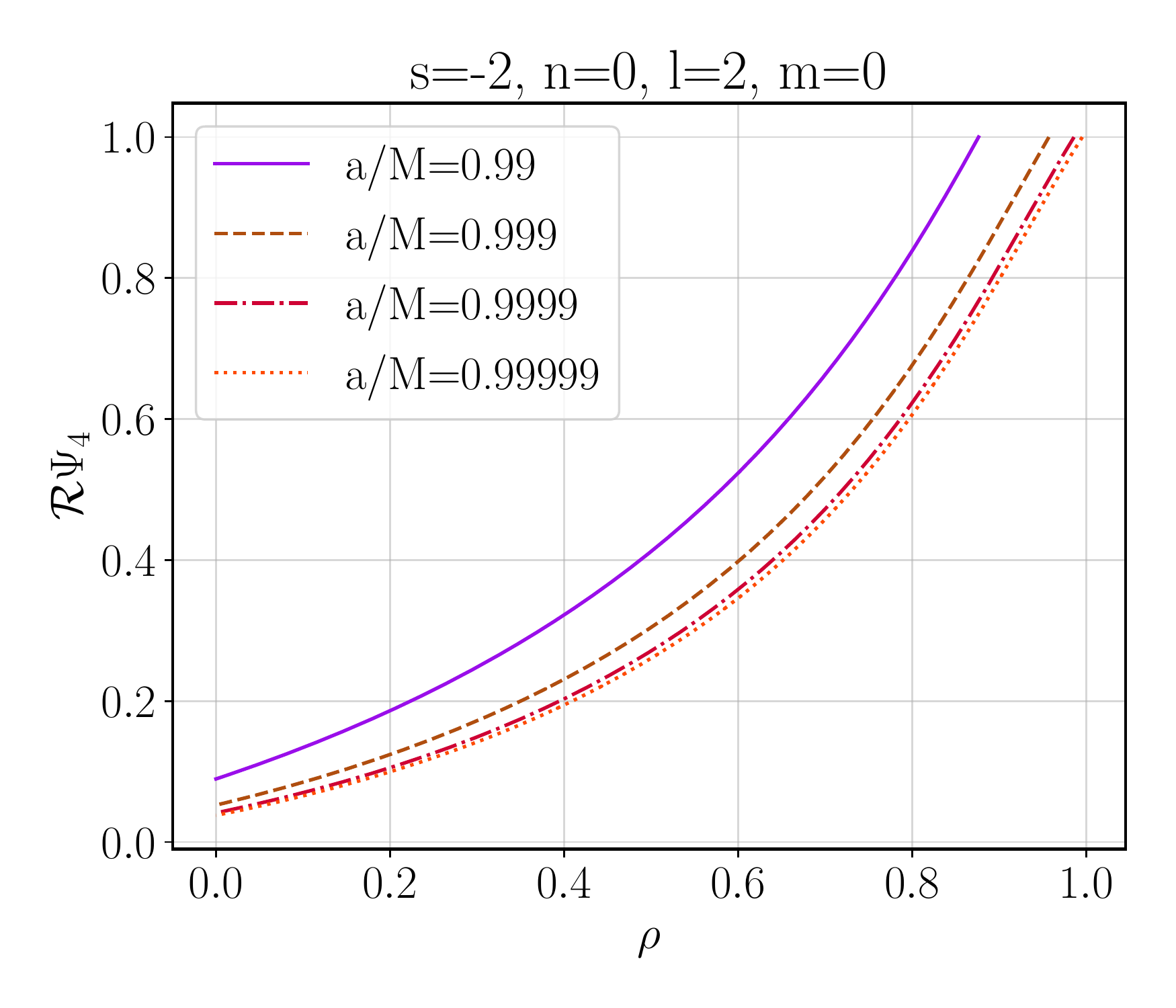}
    \includegraphics[width=0.45\columnwidth]{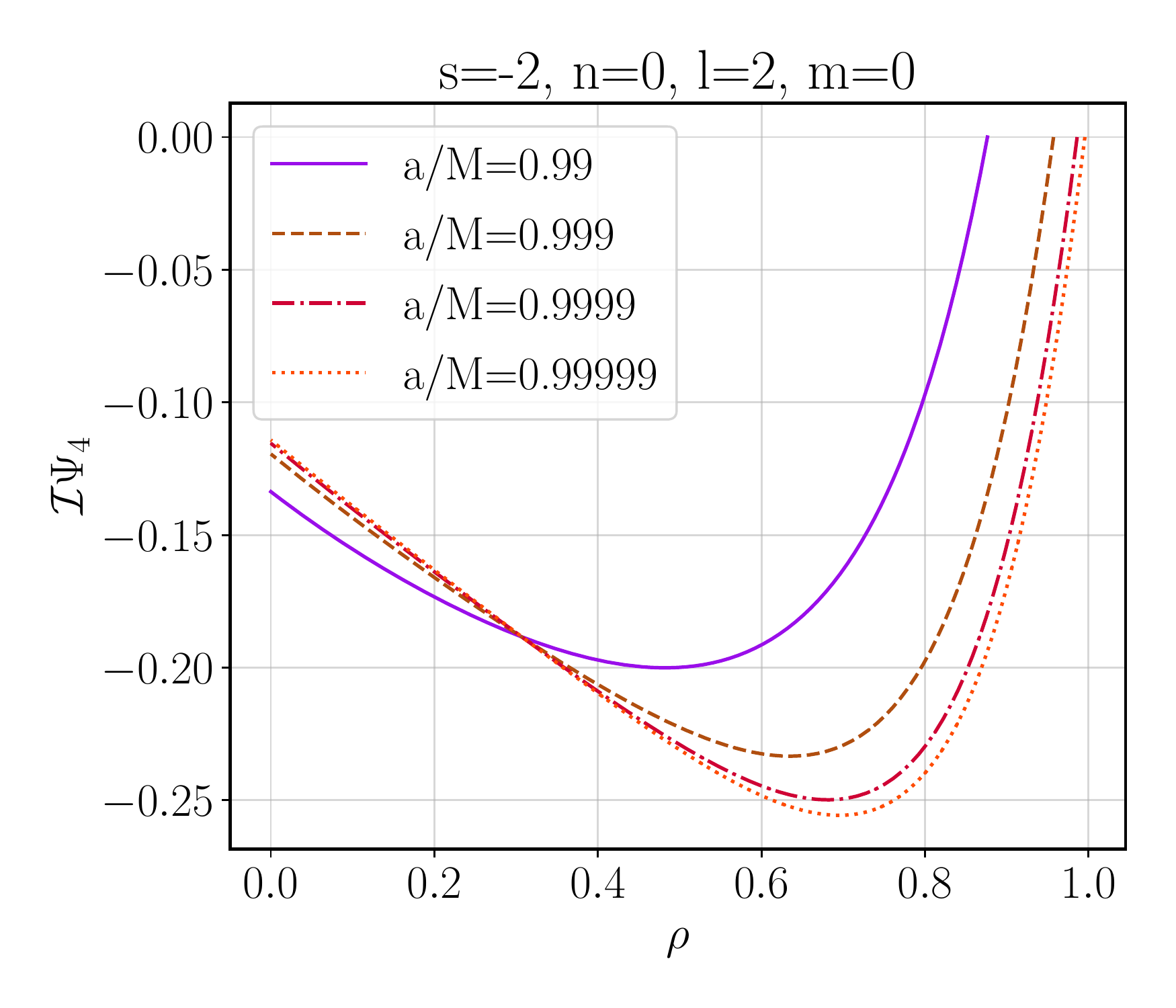}
    \includegraphics[width=0.45\columnwidth]{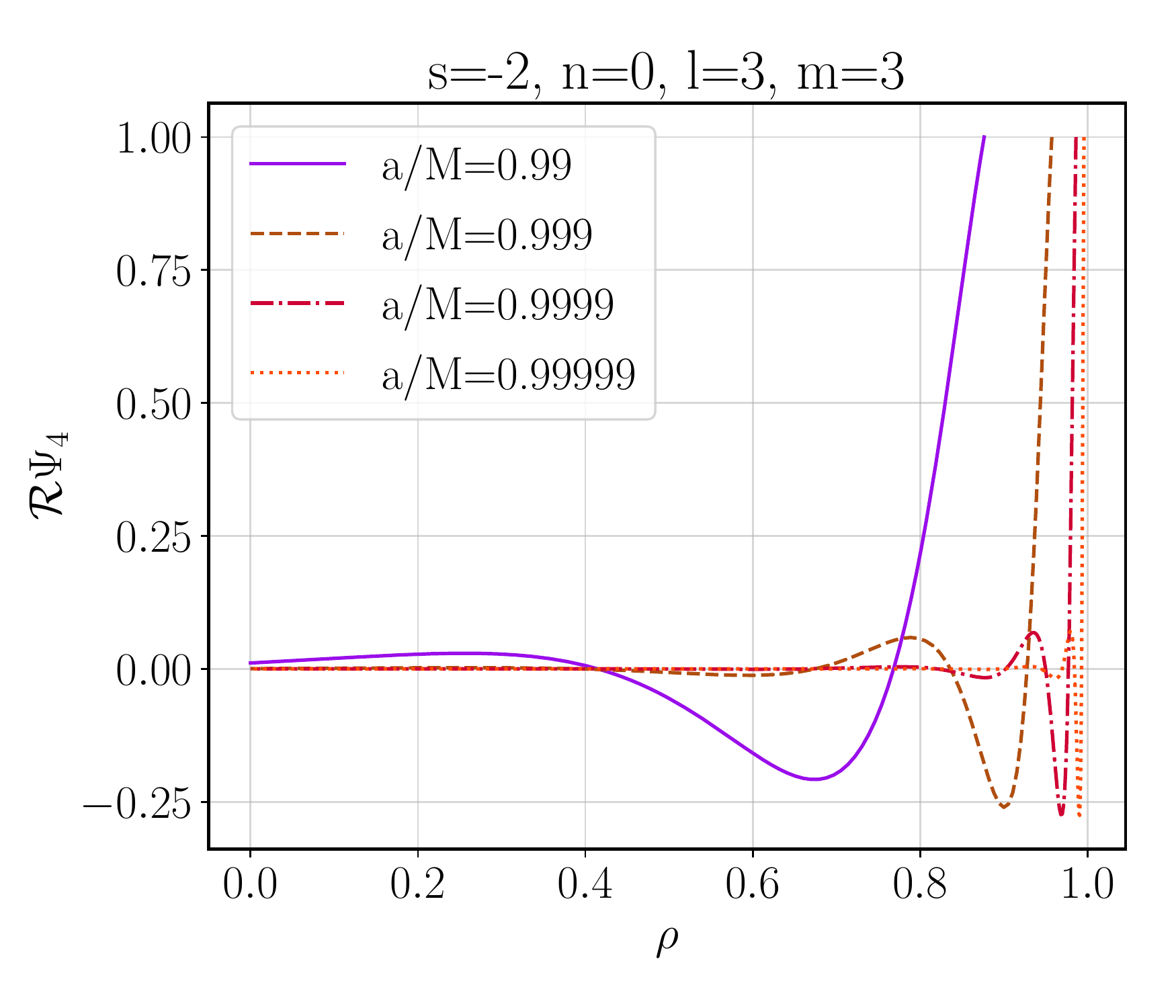}
    \includegraphics[width=0.45\columnwidth]{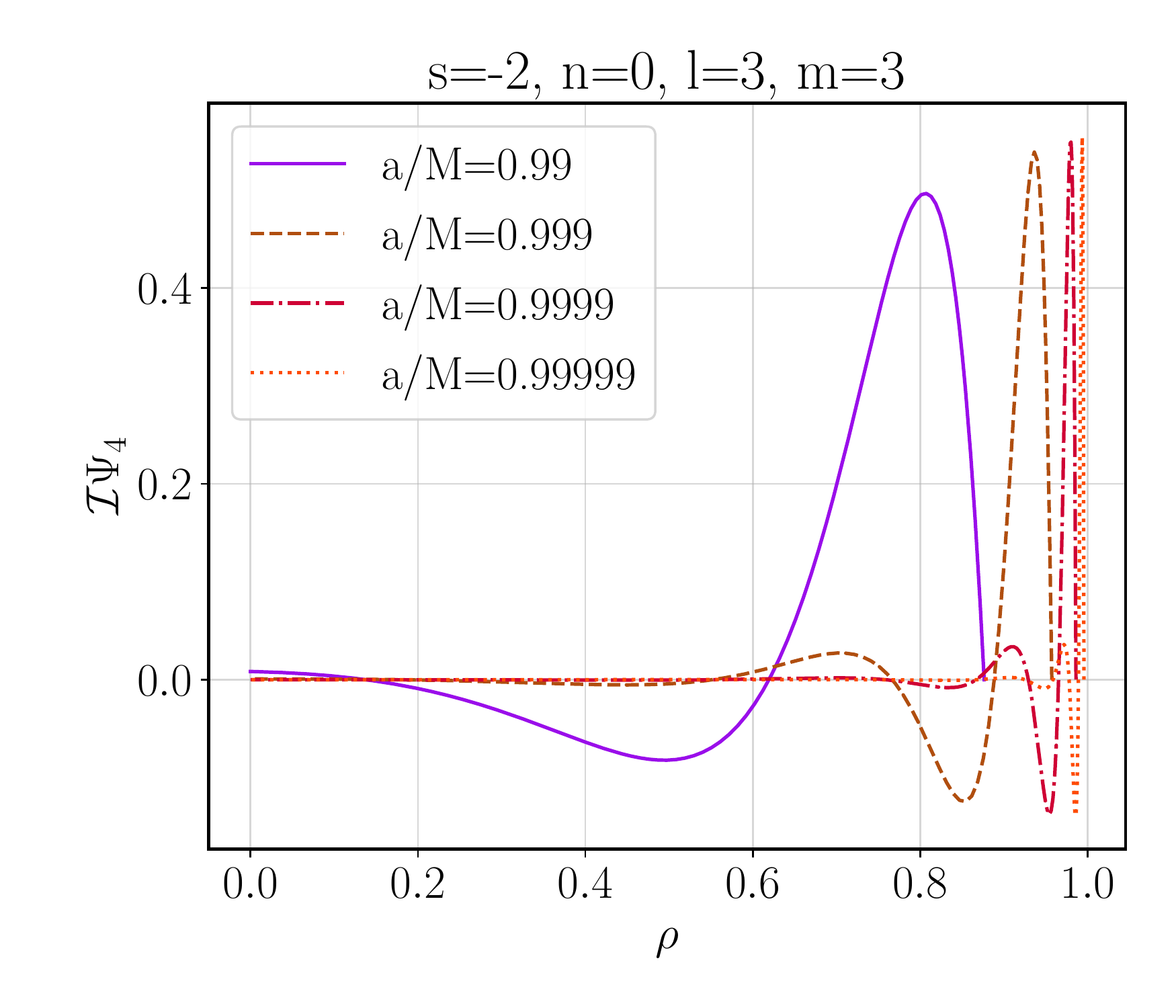}
\end{center}
   \caption{
      Other example $s=-2$ quasinormal mode radial eigenfunctions in
      the near-extremal limit. 
      Here we only plot the real and imaginary parts of $\Psi_4$, 
      and do not plot the Chebyshev or angular coefficients.
      We see that the $l=2,m=0,-2$ QNEs remain smooth
      as $a/M\to1$, which the radial derivative of the
      $n=0,l=m=3$ QNEs appear to blow 
      up in the limit $a/M\to1$.
      The $l=2,m=0,-2$ modes we consider are not zero-damped in 
      the extremal limit, while the $l=m=3$ mode we consider is 
      (see Table.~\ref{table:qnm}); for more discussion of zero-damped
      modes see \cite{Hod:2008zz,Hod:2009td,Yang:2012pj,Yang:2013uba}.
   }
\label{fig:sm2_more_radial}
\end{figure*}

There exists a family of QNMs whose imaginary part
tends to zero in the extremal black hole spin limit ($a/M\to 1$).
In \cite{Yang:2012pj,Yang:2013uba} 
these modes were called \emph{zero-damped} QNMs, 
as the imaginary part of a QNM determines its characteristic damping time
(see also \cite{Hod:2008zz,Hod:2009td}).
These modes exist for all $l,m\geq0$, and the mode
frequency takes the form 
\begin{equation}
   \label{eq:analytic_zero_damped_formula_qnm}
   M\omega 
   \approx 
   \frac{m}{2}
   - 
   \Omega_{\epsilon}\left(\frac{\epsilon}{2}\right)^{1/2}
   +
   \mathcal{O}\left(\epsilon\right)
   ,
\end{equation}
where we have defined
\begin{eqnarray}
   \Omega_{\epsilon}
   &\equiv
   \left(
      \frac{7}{4}m^2
      -
      \left(s+\frac{1}{2}\right)^2
      -
      {}_s\Lambda^m_l
   \right)^{1/2}
   +
   i \left(n+\frac{1}{2}\right)
   ,\\
   \epsilon
   &\equiv 
   1-\frac{a}{M}
   .
\end{eqnarray} 

In near-extremal limit, we observe that the QNEs associated
with the zero-damped QNMs (at least for spin-weights $s=-2,-1$) 
become sharply peaked near the black hole horizon;
see Fig.~\ref{fig:sm2_highspin} and Fig.~\ref{fig:sm1_highspin}.
We find that stably solving for the radial part of the
QNE for spins very near extremality
requires the use the use of higher-precision arithematic.
We present the largest spin ($a=0.99999$)
QNMs and QNEs that we can resolve with
$244$ radial Chebyshev coefficients and $1024$ bits of floating-point precision.
As noted by Cook and Zalutskiy \cite{Cook:2014cta}, 
we do not need increasingly more
spin-weighted spherical harmonics to resolve the angular part of
the QNE as we approach the extremal black hole limit.
We show three more examples near-extremal $s=-2$ QNEs 
in Fig.~\ref{fig:sm2_more_radial}.
From those figures, we see that the $n=0,l=2,m=0,-2$ QNEs 
appear to remain smooth in the extremal limit, while the radial derivative
of the $n=0,l=m=3$ QNEs appear to blow up in that limit.
We note that the $n=0,l=2,m=0,-2$ QNMs are not zero-damped, 
while the $n=0,l=m=3$ QNMs are; see
Table~\ref{table:qnm} or \cite{Yang:2012pj} for more discussion.

We provide a semi-analytic argument
to explain growth in the radial gradient of the zero-damped QNE 
grow near the black hole horizon as $\epsilon\to0$.
The derivative of the
near-horizon solution (Eq.~\ref{eq:near_horizon_solution}), 
to leading order in $x\ll1$ is:
\begin{equation}
   \label{eq:expansion_derivative}
   \frac{dR_{\mathcal{H}}}{d\rho}
   =
   \mathcal{A}_{\mathcal{H}}\frac{1}{\rho_+}\left(
      \frac{a_{\mathcal{H}}b_{\mathcal{H}}}{c_{\mathcal{H}}}
      +
      \mathcal{O}\left(x\right)
   \right)
   .
\end{equation}
We next expand in the near-extremal limit (i.e. in $\epsilon$), 
and plug in the zero-damped value for the QNM 
(Eq.~\ref{eq:analytic_zero_damped_formula_qnm}).
This formula fits our numerical data for the zero-damped modes
well when $\epsilon\ll1$; see Table~\ref{table:qnm}.
Plugging into Eq.~\ref{eq:expansion_derivative} the above value for $\omega$,
along with using Eqns.~\ref{eq:definitions_near_horizon_1},
\ref{eq:definitions_near_horizon_3}, and
\begin{equation}
   a/M
   =
   1
   +
   \epsilon
   ,\qquad
   M\rho_+
   =
   1
   -
   \left(2\epsilon\right)^{1/2}
   +
   \mathcal{O}\left(\epsilon\right)
   ,
\end{equation}
we find that
\begin{eqnarray}
   \frac{dR_{\mathcal{H}}}{d\rho}
   =
   \frac{\mathcal{A}_{\mathcal{H}}M}{2\sqrt{2}}\times
   \left(
      \frac{
         \frac{3}{4}m^2
         -
         2s
         +
         im\left(2s-1\right)
         +
         {}_s\Lambda^m_l
      }{
         1
         -
         s
         -
         im
         +
         i\Omega_{\epsilon}
      }
      \frac{1}{\sqrt{\epsilon}}
      +
      \mathcal{O}\left(1\right)
   \right)
   .
\end{eqnarray}
We see that as $\epsilon\to0$ that the derivative of the near-horizon 
($x\ll1$) solution blows up when we plug in the zero-damped mode ansatz.
This calculation agrees qualitatively with what we see in our numerical
computations of the zero-damped QNEs;
(see Figs.~\ref{fig:sm2_highspin},
\ref{fig:sm1_highspin},
\ref{fig:sm2_more_radial}).

In the coordinates we are using, the
radial proper distance does not blow up near the black hole horizon;
see Eq.~\ref{eq:kerr_radial_proper_distance}.
Because of this, we can conclude that 
$  \frac{dR_{\mathcal{H}}}{dr}
   =
   -
   \frac{1}{r^2}\frac{dR_{\mathcal{H}}}{d\rho}
   ,
$
also grows larger in the limit $a/M\to1$
near the black hole horizon for the zero-damped QNMs.
%%%%%%%%%%%%%%%%%%%%%%%%%%%%%%%%%%%%%%%%%%%%%%%%%%%%%%%%%%%%%%%%%%%%%%%%%%%%%%
\section{Discussion\label{Discussion}}
   We have computed several of the quasinormal modes
and their associated quasinormal eigenfunctions
of the Teukolsky equation.
These calculations were performed in horizon-penetrating, 
hyperboloidally-compactified (HPHC) coordinates.
With these coordinates (and with a suitable choice of tetrad), 
the QNEs of the Teukolsky equation are regular, including at
the black hole horizon and future null infinity.
In the process of computing the QNMs and QNEs,
we have found that the eigenfunctions for the zero-damped modes
develop a steep radial gradient near the black hole horizon
in the near-extremal Kerr limit ($a/M\to1$).
This feature of the mode solutions makes resolving the zero-damped QNEs 
increasingly difficult as the black hole spin 
approaches extremality.

For future work, 
it would be interesting to further investigate the properties
of QNE solutions, such as the properties
of the solutions in the limit of large overtone number.
The properties of the QNE solutions
in the near-extremal Kerr limit also deserve further study, 
given the variety QNM solutions 
that can be found in that limit 
\cite{Hod:2008zz,Hod:2009td,Yang:2012pj,Yang:2012he,Yang:2013uba}.
Aretakis has shown that extremal Kerr black hole spacetimes are
unstable at the black hole horizon \cite{Aretakis:2011gz,Aretakis:2012ei}. 
This instability arises from outgoing wave solutions
that have support
at the black hole horizon, 
which develop gradients in the radial direction 
that grow unbounded over time. 
Similar results have been found for the zero-damped QNM solutions of
the near-extremal Kerr spacetime 
\cite{Gralla:2017lto,Gralla:2018xzo}.
It would be interesting to connect the results of our study to
previous work on the Aretakis instability (for example, one may
expect that given Aretakis' result,
zero-damped QNE solutions in the extremal limit become
step-function like).
As we mentioned in the Introduction, it would be interesting to apply
our results to a computation of the pseudospectrum of quasinormal modes
of the Teukolsky equation, thus extending the work of 
\cite{Jaramillo:2020tuu,Destounis:2021lum} to that spacetime. 
Finally, another direction for potential work would be 
to extend this work to computing the QNEs to the
Kerr-Newman spacetime in HPHC coordinates \cite{Dias:2015wqa}. 

The numerical methodology we use to compute the QNEs 
could be improved as well.
While the pseudospectral Chebyshev
approach we use to compute the radial QNEs 
naturally impose regular boundary conditions on the solution,
we have found that we had to use many ($N_{(\rho)}>200$) 
Chebyshev coefficients to properly resolve the zero-damped
QNEs in the near-extremal limit. 
We have also found that significantly more 
Chebyshev coefficients are needed to resolve higher overtones,
regardless of the value of the black hole spin.
As Chebyshev derivative matrices are very poorly conditioned, this
then requires the use of higher-precision arithematic, which dramatically
slows down the speed at which we can compute quasinormal mode solutions.
A modification of Leaver's method (that is, a spectral expansion
in a series of rational polynomials of $\rho$) may provide a more
reliable and stable method to obtain quasinormal modes in HPHC coordinates.
%%%%%%%%%%%%%%%%%%%%%%%%%%%%%%%%%%%%%%%%%%%%%%%%%%%%%%%%%%%%%%%%%%%%%%%%%%%%%%
\ack
 I am grateful to 
 Emanuele Berti,
 Kyriakos Destounis,
 and
 Rodrigo Panosso Macedo
 for their detailed comments on an earlier draft of this note,
 and to Alejandro Cardenas-Avendano,
 Alex Lupsasca,
 An\i{}l Zengino\u{g}lu, 
 Elena Giorgi, 
 Frans Pretorius,
 and Nicholas Loutrel
 for discussions that motivated this project.
 During the course of this work I was
 supported by STFC Research Grant No. ST/V005669/1.
 The simulations presented in this paper made use of the Cambridge
 Service for Data Driven Discovery (CSD3), part of
 which is operated by the University of Cambridge
 Research Computing on behalf of the STFC DiRAC
 HPC Facility (www.dirac.ac.uk).
 %%%%%%%%%%%%%%%%%%%%%%%%%%%%%%%%%%%%%%%%%%%%%%%%%%%%%%%%%%%%%%%%%%%%%%%%%%%%%%
\appendix
%%%%%%%%%%%%%%%%%%%%%%%%%%%%%%%%%%%%%%%%%%%%%%%%%%%%%%%%%%%%%%%%%%%%%%%%%%%%%%
\subsection{Quasinormal modes and convergence
   \label{sec:quasinormal_modes_and_convergence}
}
In Table.~\ref{table:qnm}
we list QNMs as computed using the code \cite{code_online}.
These QNMs agree with those computed from the \texttt{qnm} 
code \cite{Stein:2019mop} to the precision given in the table
(except for the $s=-2,n=0,l=m=-2$ mode; see the caption to the Table).
\begin{table}[H]
\begin{center}
\begin{tabular}{ |c c c c| } 
   \hline
   (s,n,l,m) & a & $M\omega$ & ${}_{s}\Lambda^m_l$ 
   \\ \hline \hline 
   $(-2,0,2,2)$  & $0$       & $0.3736716-0.0889623i$ & $4$ \\
   $(-2,0,2,2)$  & $0.5$     & $0.4641230-0.0856388i$ & $3.3423+0.1292i$ \\
   $(-2,0,2,2)$  & $0.7$     & $0.5326002-0.0807928i$ & $2.9032+0.1832i$ \\
   $(-2,0,2,2)$  & $0.9$     & $0.6716142-0.0648692i$ & $2.1098+0.2111i$ \\
   $(-2,0,2,2)$  & $0.99$    & $0.8708926-0.0293904i$ & $1.1196+0.1180i$ \\
   $(-2,0,2,2)$  & $0.999$   & $0.9558544-0.0105305i$ & $0.7357+0.0443i$ \\
   $(-2,0,2,2)$  & $0.9999$  & $0.9856735-0.0034686i$ & $0.6055+0.0148i$ \\
   $(-2,0,2,2)$  & $0.99999$ & $0.9954317-0.0011112i$ & $0.5633+0.0047i$ \\
   \hline 
   $(-2,0,2,-2)$ & $0.99$    & $0.2921067-0.0880523i$ & $4.7163-0.1966i$ \\
   $(-2,0,2,-2)$ & $0.999$   & $0.2916086-0.0880285i$ & $4.7212-0.1981i$ \\
   $(-2,0,2,-2)$ & $0.9999$  & $0.2915590-0.0880261i$ & $4.7217-0.1983i$ \\
   $(-2,0,2,-2)$ & $0.99999$ & $0.2915567-0.0880160i$ & $4.7218-0.1981i$ \\ 
   \hline 
   $(-2,0,2,0)$  & $0.99$    & $0.4236846-0.0727008i$ & $3.9102+0.0319i$ \\
   $(-2,0,2,0)$  & $0.999$   & $0.4249978-0.0718986i$ & $3.9079+0.0322i$ \\
   $(-2,0,2,0)$  & $0.9999$  & $0.4251304-0.0718155i$ & $3.9077+0.0323i$ \\
   $(-2,0,2,0)$  & $0.99999$ & $0.4251435-0.0718072i$ & $3.9076+0.0323i$ \\ 
   \hline 
   $(-2,0,3,3)$  & $0.99$    & $1.3230831-0.029403i$ & $6.4040+0.1043i$ \\
   $(-2,0,3,3)$  & $0.999$   & $1.4397481-0.010530i$ & $5.9312+0.0397i$ \\
   $(-2,0,3,3)$  & $0.9999$  & $1.4804730-0.003469i$ & $5.7714+0.0133i$ \\
   $(-2,0,3,3)$  & $0.99999$ & $1.4937761-0.001111i$ & $5.7197+0.0043i$ \\ 
   \hline 
   $(-1,0,1,1)$  & $0.0$     & $0.2482633-0.092488i$ & $2$ \\
   $(-1,0,1,1)$  & $0.5$     & $0.2940910-0.087677i$ & $1.8419+0.0511i$ \\
   $(-1,0,1,1)$  & $0.7$     & $0.3266554-0.081869i$ & $1.7436+0.0724i$ \\
   $(-1,0,1,1)$  & $0.9$     & $0.3875811-0.065625i$ & $1.5832+0.0835i$ \\
   $(-1,0,1,1)$  & $0.99$    & $0.4633988-0.031292i$ & $1.4185+0.0482i$ \\
   $(-1,0,1,1)$  & $0.999$   & $0.4896711-0.011609i$ & $1.3700+0.0185i$ \\
   $(-1,0,1,1)$  & $0.9999$  & $0.4971357-0.003805i$ & $1.3573+0.0061i$ \\
   $(-1,0,1,1)$  & $0.99999$ & $0.4991753-0.001177i$ & $1.3539+0.0019i$ \\
   \hline
\end{tabular}
\caption{Computed QNMs and separation constants, computed using the 
      code \cite{code_online},
      which implements the algorithm in Sec.~\ref{sec:discretization_method}.
      To produce accurate results at high spin using the radial pseudospectral
      method, we had to resort to using
      higher-precision arithematic (here: 1024 bits of precision).
      These results agree with those computed from the \texttt{qnm} 
      code \cite{Stein:2019mop} to the precision presented, except
      for the $s=-2,n=0,l=2,m=-2$ mode, where there is a difference
      between the qnm code in the last two decimal places for the values
      of $\omega$ and ${}_s\Lambda^m_l$.
   }
   \label{table:qnm} 
\end{center}
\end{table}

\begin{figure*}[h]
\begin{center}
    \includegraphics[width=0.45\columnwidth]{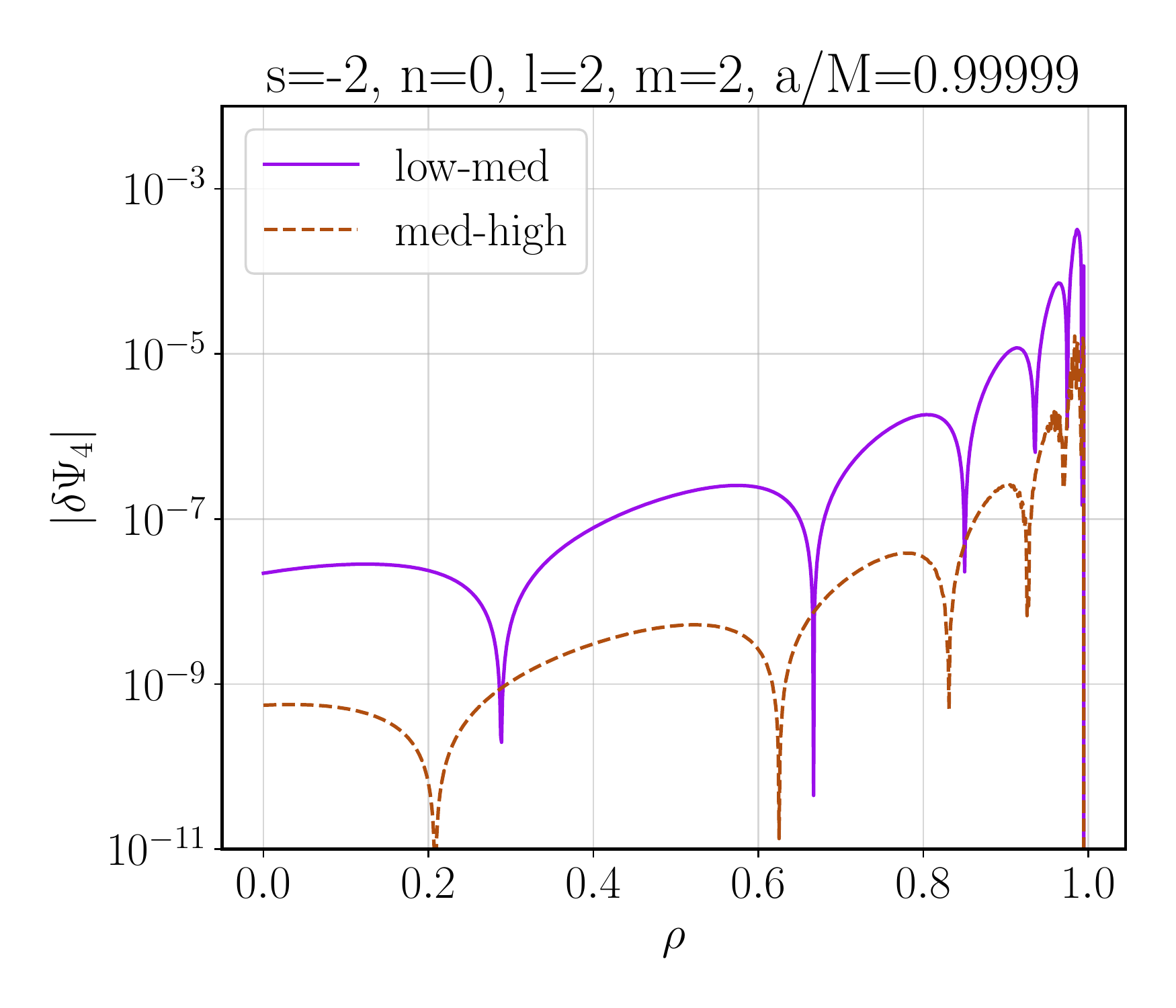}
\end{center}
   \caption{
      Convergence study of the radial part of the $s=-2,n=0,l=2,m=2$
      QNE for a black hole spin parameter $a/M=0.99999$.
      These calculations were performed
      with $1024$ bit precision, and the resolutions of the
      low, med, high resolution calculations were respectively
      ($N_{(\rho)}=204,N_{(\theta)}=20$),
      ($N_{(\rho)}=224,N_{(\theta)}=22$),
      and
      ($N_{(\rho)}=244,N_{(\theta)}=24$).
   }
\label{fig:quasinormal_mode_convergence}
\end{figure*}

We next present an example convergence test of a QNE 
in Fig.~\ref{fig:quasinormal_mode_convergence}.
We see that the pointwise difference between a ``low'' and ``medium'' resolution
calculations is larger than the difference between a ``medium''
and ``high'' resolution calculation. These calculations were performed
with $1024$ bit precision, and the resolutions of the
low, med, high resolution calculations were respectively
$(N_{(\rho)}=204,N_{(\theta)}=20)$,
$(N_{(\rho)}=224,N_{(\theta)}=22)$,
and
$(N_{(\rho)}=244,N_{(\theta)}=24)$.

%%%%%%%%%%%%%%%%%%%%%%%%%%%%%%%%%%%%%%%%%%%%%%%%%%%%%%%%%%%%%%%%%%%%%%%%%%%%%%
\section{
   Some properties of the Jacobi polynomials and 
   spin-weighted spherical harmonics 
   \label{sec:orthogonal_polynomials}
}
%%%%%%%%%%%%%%%%%%%%%%%%%%%%%%%%%%%%%%%%%%%%%%%%%%%%%%%%%%%%%%%%%%%%%%%%%%%%%%
\subsection{Jacobi polynomials
   \label{sec:jacobi_polynomials}
}
Our notation follows \cite{beals2016special} (see also, e.g.
\cite{NIST:DLMF} for a more general reference).
The Jacobi polynomials 
are orthogonal with respect to the weight 
$w=\left(1-x\right)^{\alpha}\left(1+x\right)^{\beta}$
on the interval $(-1,1)$. They are denoted by: 
\begin{equation}
   P^{(\alpha,\beta)}_n(x)
   ,\qquad
   n=0,1,2,...
   \alpha,\beta>-1
\end{equation}
The Jacobi polynomials satisfy the following orthogonality condition
\begin{eqnarray}
   \fl
   \int_{-1}^1dx 
      \left(1-x\right)^{\alpha}\left(1+x\right)^{\beta}
      P_n^{(\alpha,\beta)}(x)
      P_m^{(\alpha,\beta)}(x)
   =
   \nonumber\\
   \frac{
      2^{\alpha+\beta+1}
   }{
      2n+\alpha+\beta+1
   }
   \frac{
      \Gamma\left(n+\alpha+1\right)
      \Gamma\left(n+\beta+1\right)
   }{
      n!\Gamma\left(n+\alpha+\beta+1\right)
   }
   \delta_{nm}
   .
\end{eqnarray}
The derivative and recursion relations for the Jacobi polynomials are
\begin{eqnarray}
   \label{eq:jacobi_derivative}
   \fl
   \frac{d}{dx}P^{(\alpha,\beta)}_n(x)
   &=
   \frac{1}{2}\left(n+\alpha+\beta+1\right)
   P^{(\alpha+1,\beta+1)}_{n-1}(x)
   ,\\
   \fl
   \frac{(2n+2)(n+\alpha+\beta+1)}{2n+\alpha+\beta+1}
   \label{eq:jacobi_recursion}
   P^{(\alpha,\beta)}_{n+1}(x)
   &=
   \left[
      \frac{\alpha^2-\beta^2}{2n+\alpha+\beta}
      +
      \left(2n+\alpha+\beta+2\right)x
   \right]
   P^{(\alpha,\beta)}_{n}(x)
   \nonumber\\
   &
   -
   \frac{
      2(2n+\alpha+\beta+2)(n+\alpha)(n+\beta)
   }{
      (2n+\alpha+\beta)(2n+\alpha+\beta+1)
   }
   P^{(\alpha,\beta)}_{n-1}(x)
   .
\end{eqnarray}
%%%%%%%%%%%%%%%%%%%%%%%%%%%%%%%%%%%%%%%%%%%%%%%%%%%%%%%%%%%%%%%%%%%%%%%%%%%%%%
\subsection{
   Spin-weighted spherical harmonics
   \label{sec:properties_swal}
}
We work in spherical polar coordinates.
The spin-weighted spherical harmonics are a complete set of
orthogonal functions on the
sphere for spin-weighted functions.
They are eigenfunctions for the spin-weighted spherical Laplacian:
\begin{eqnarray}
\label{eq:spin_weighted_associated_legendre_eom}
   \fl
   \frac{1}{\sin\theta}\partial_{\theta}\left(
      \sin\theta\partial_{\theta}
      {}_sY^m_l\left(\theta,\phi\right)
   \right)
   \nonumber\\
   +
   \left(
      s
      -
      \frac{\left(-i\partial_{\phi}+s\cos\theta\right)^2}{\sin^2\theta}
      +
      \left(l-s\right)\left(l+s+1\right)
   \right)
   {}_sY^m_l\left(\theta,\phi\right)
   =
   0
   .
\end{eqnarray}
The spin-weighted spherical harmonics can be written as
\begin{equation}
   {}_sY^m_l(\theta,\phi)
   =
   e^{im\phi}
   {}_sP^m_l\left(\theta\right)
   ,
\end{equation}
where ${}_sP^m_l(y)$ is the spin-weighted associated Legendre polynomial.
These functions are related to the Jacobi polynomials
via (e.g. \cite{2018arXiv180410320V,Ripley:2020xby}):
\begin{equation}
   {}_sP^m_l(y)
   =
   {}_s\mathcal{N}^m_l
   \left(1-y\right)^{\alpha}
   \left(1+y\right)^{\beta}
   P^{(\alpha,\beta)}_{n}(y)
   ,
\end{equation}
where
$y
   \equiv
   -
   \cos\theta
$,
$\alpha
   \equiv
   \left|m-s\right|
   ,
   \beta
   \equiv
   \left|m+s\right|
   ,
   n
   \equiv
   l
   -
   \frac{\alpha+\beta}{2}
$, 
and
\begin{equation}
   \label{eq:normalization_swal}
   {}_s\mathcal{N}^m_l
   \equiv
   \left(-1\right)^{\mathrm{max}(m,-s)}
   \left(
      \frac{
         2n+\alpha+\beta+1
      }{
         2^{\alpha+\beta+1}
      }
      \frac{
         n!\left(n+\alpha+\beta\right)!
      }{
         \left(n+\alpha\right)!
         \left(n+\beta\right)!
      }
   \right)^{1/2}
   .
\end{equation}

From the recursion relation for the Jacobi polynomials,
Eq.~\ref{eq:jacobi_recursion}, and the normalization of
the spin-weighted spherical harmonics, Eq.~\ref{eq:normalization_swal},
we have \cite{1977RSPSA.358...71B,Cook:2014cta}
\begin{equation}
   \label{eq:yslm_relations}
   y{}_sP^m_l(y)
   =
   {}_s\mathcal{A}^m_l {}_sP^m_{l+1}(y)
   +
   {}_s \mathcal{B}^m_l {}_sP^m_{l}(y)
   +
   {}_s \mathcal{C}^m_l {}_sP^m_{l-1}(y)
   ,
\end{equation}
where
\begin{eqnarray}
   \fl
   {}_s\mathcal{A}^m_l
   &=
   \frac{
      2
   }{
      (2n+\alpha+\beta+2)
   }
   \left[
      \frac{
         (n+1)
         (n+\alpha+1)
         (n+\beta+1)
         (n+\alpha+\beta+1)
      }{
         (2n+\alpha+\beta+1)
         (2n+\alpha+\beta+3)
      }
   \right]^{1/2}
   \nonumber \\
   \fl
   &=
   \left[
      \frac{
         \left((l+1)^2 - s^2\right)
         \left((l+1)^2 - m^2\right)
      }{
         (l+1)^2
         (2l+1)
         (2l+3)
      }
   \right]^{1/2}
   ,\\
   \fl
   {}_s\mathcal{B}^m_l
   &=
   -
   \frac{
      \alpha^2-\beta^2
   }{
      (2n+\alpha+\beta)
      (2n+\alpha+\beta+2)
   }
   \nonumber\\
   \fl
   &=
   -
   \frac{
      ms
   }{
      l(l+1)
   }
   ,\\
   \fl
   {}_s\mathcal{C}^m_l
   &=
   \frac{
      2
   }{
      (2n+\alpha+\beta)
   }
   \left[
      \frac{
         n(n+\alpha)(n+\beta)(n+\alpha+\beta)
      }{
         (2n+\alpha+\beta+1)(2n+\alpha+\beta-1)
      }
   \right]^{1/2}
   \nonumber\\
   \fl
   &=
   \left[
      \frac{
         \left(l^2-s^2\right)
         \left(l^2-m^2\right)
      }{
         l^2\left(4l^2-1\right)
      }
   \right]^{1/2}
   .
\end{eqnarray}
%%%%%%%%%%%%%%%%%%%%%%%%%%%%%%%%%%%%%%%%%%%%%%%%%%%%%%%%%%%%%%%%%%%%%%%%%%%%%%
\section{
   Converting angular ODE to a sparse matrix equation in spectral space 
   \label{sec:angular_ODE_sparse}
}
For completeness we review a spectral method that converts
the angular equation, Eq.~\ref{eq:spheroidal_laplacian},
into a sparse linear-algebra problem\cite{Cook:2014cta,Hughes:1999bq}.
We recall that Eq.~\ref{eq:spheroidal_laplacian}
is solved by the spin-weighted spheroidal harmonics
\begin{equation}
   S
   =
   {}_sS^m_l\left(\theta,c\right)
   .
\end{equation}
When $a=0$ the equation reduces to that of the 
spin-weighted spherical harmonics.
We expand the spin-weighted spheroidal harmonics in terms of
spin-weighted spherical harmonics (which form a basis
for spin-weighted functions on the sphere):
\begin{equation}
   {}_sS^m_l\left(\theta,\phi,c\right)
   =
   \sum_{l'} 
   {}_sg^m_{l,l'}\left(a\omega\right){}_sY^m_{l'}\left(\theta,\phi\right)
   =
   \sum_{l'} 
   {}_sg^m_{l,l'}\left(a\omega\right){}_sP^m_{l'}\left(\theta\right)e^{im\phi}
   ,
\end{equation}
where ${}_sP^m_{l'}$ are the spin-weighted associated Legendre polynomials.
This approach was first used in the context of numerically computing
QNM in \cite{Cook:2014cta} (although similar earlier semi-analytic 
works include \cite{1973ApJ...185..649P,1977RSPSA.358...71B}).
We expand $S$ in terms of spherical harmonics, 
and use Eq.~\ref{eq:spin_weighted_associated_legendre_eom}
to rewrite Eq.~\ref{eq:spheroidal_laplacian} as
\begin{eqnarray}
   \fl
   &
   \frac{1}{\sin\theta}
   \frac{d}{d\theta}\left(\sin\theta\frac{d{}_sS^m_l}{d\theta}\right)
   +
   \left(
      s
      -
      \frac{\left(m+s\cos\theta\right)^2}{\sin^2\theta}
      -
      2a\omega s\cos\theta
      +
      a^2\omega^2\cos^2\theta
      +
      {}_s\Lambda^m_l
   \right)
   {}_sS^m_l
   \nonumber\\
   \fl
   &
   =
   e^{im\phi}
   \sum_{l'}
   \left(
      -
      2a\omega s\cos\theta
      +
      a^2\omega^2\cos^2\theta
      -
      \left(l'-s\right)\left(l'+s+1\right)
      +
      {}_s\Lambda^m_l
   \right){}_sg^m_{l,l'}{}_sP^m_{l'}\left(\theta\right)
   =
   0
   \nonumber\\
   \fl
   &
   = 
   e^{im\phi}
   \left(
      \vec{{}_sg_l^m}
      ,
      \left[
         -
         \hat{M}_{(\theta)}^T
         +
         {}_s\Lambda^m
         \hat{I}
   \right]
   \vec{{}_sP^m}
   \right)
   ,
\end{eqnarray}
where
\begin{equation}
   \left[\hat{M}_{\theta}^T\right]_{l,l'}
   \equiv
   \left(a\omega\right)^2\left[\hat{Y}_1^2\right]_{l,l'}
   -
   2s\left(a\omega\right)\left[\hat{Y}_1\right]_{l,l'}
   -
   \delta_{l,l'}
   \left(l'-s\right)
   \left(l'+s+1\right)
   .
\end{equation}
Here we have defined the matrix
(see Eq.~\ref{eq:yslm_relations})
\begin{equation}
   y\vec{{}_sP^m}
   \equiv
   \hat{Y}_1\vec{{}_sP^m}
   .
\end{equation}
For a fixed $(s,m,a)$, we then have the sparse matrix eigenvalue equation 
\begin{equation}
   \left(
      \hat{M}_{(\theta)}
      -
      \Lambda \hat{I}
   \right)
   \vec{g}
   =
   0
   .
\end{equation}
%%%%%%%%%%%%%%%%%%%%%%%%%%%%%%%%%%%%%%%%%%%%%%%%%%%%%%%%%%%%%%%%%%%%%%%%%%%%%%
\section{
   Chebyshev pseudospectral discretization of the radial ODE 
   \label{sec:radial_ODE_pseudospectral}
}
We briefly review Chebyshev (pseudospectral) collocation methods
\cite{trefethen2000spectral,boyd2001chebyshev}.
See also \cite{Jansen:2017oag,Jaramillo:2020tuu} for other recent
examples of applying pseudospectral methods to calculating the
QNMs and QNEs of the various wave equations for
spherically symmetric black hole spacetimes.

The Chebyshev polynomials (of the first kind) of order $n$ are given by
\begin{equation}
   T_n\left(x\right)
   =
   \cos\left(n \arccos x\right)
   ,\qquad
   x\in\left[-1,1\right]
   .
\end{equation}
The Chebyshev polynomials form an orthonormal basis for functions 
$f\in L^2\left([-1,1],w(x)dx\right)$, 
where $w(x) = \left(1-x^2\right)^{-1/2}$, in particular they satisfy 
\begin{equation}
   \int_{-1}^1\frac{dx}{\sqrt{1-x^2}}
      T_n\left(x\right)
      T_m\left(x\right)
   =
   \left\{\begin{array}{r@{}l@{\qquad}l}
      0 & & n\neq m \\[\jot]
      \frac{\pi}{2} & & n=m\neq0 ,\\
      \pi & & n=m=0 
   \end{array}\right.
\end{equation}
One can expand essentially any sufficiently smooth function on the 
interval $[-1,1]$ as a sum of Chebyshev polynomials
\begin{equation}
   f\left(x\right)
   =
   \frac{1}{2}c_0
   +
   \sum_{n=1}^{\infty} c_n T_n(x)
   .
\end{equation}

We discretize the radial ODE using the
Chebyshev extreme points as the collocation points.
On the interval $[-1,1]$, they are
\begin{equation}
   x_j = \cos\left(\frac{\pi j}{N}\right)
   ,\qquad
   j=0,1,...,N
   ,
\end{equation}
which we map to the interval $[0,\rho_+]$ via 
$\rho_j = \rho_+\left(x_j+1\right)/2$.
With this method, the derivative operators are converted into dense
matrices, whose components take the form
\begin{equation}
   D_{ij}
   =
   \frac{dT_i}{dx}\Big|_{x=x_j}
   .
\end{equation}
Explicitly we have
\begin{equation}
   D_{ij}
   =
   \left\{\begin{array}{r@{}l@{\qquad}l}
      - \frac{2N^2+1}{6}       & & i=j=N  \\[\jot]
        \frac{2N^2+1}{6}       & & i=j=0 \\
      - \frac{x_j}{2(1-x_i^2)} & & 0<i=j<N \\
        \frac{c_i}{c_j}\frac{(-1)^{i+j}}{(x_i-x_j)} & & 
         i\neq j, \qquad i,j=1,...,N-1
   \end{array}\right.
   ,
\end{equation}
where $c_0=c_N=2$ and $c_1=\cdots=c_{N-1}=1$.
We can define a second derivative matrix using
\begin{equation}
   D^{(2)}_{ij}
   =
   \sum_kD_{ik}D_{kj}
   .
\end{equation}
These derivative matrices are generally ill-conditioned
(the condition number for $D^{(2)}_{ij}$ goes as $N^4$,
where $N$ is the size of the matrix 
\cite{trefethen2000spectral,boyd2001chebyshev}),
but we have empirically found that the \texttt{eigen} 
function in the \texttt{Julia} standard library \cite{bezanson2017julia},
(sometimes augmented with higher precision arithematic using the
\texttt{GenericSchur} library \cite{generic_schur}--with floating point
numbers the \texttt{eigen} library simply calls a 
LAPACK routine \cite{lapack_book})
that we can stably obtain the eigenvalues and
eigenvectors to the radial ODE, even with matrices larger
than $100\times100$. 
%%%%%%%%%%%%%%%%%%%%%%%%%%%%%%%%%%%%%%%%%%%%%%%%%%%%%%%%%%%%%%%%%%%%%%%%%%%%%%
\bibliographystyle{iopart-num}
\bibliography{thebib}
\end{document}